\documentclass[useAMS,usenatbib]{mn2e}
\pdfoutput=1

\newcommand\Msol{$M_{\odot}$}


\usepackage{times}
\usepackage[pdftex]{graphicx}
\usepackage{aas_macros}
\usepackage{amssymb}
\usepackage{natbib}
\usepackage{multirow}
\usepackage{xspace}
\usepackage{pdflscape}
\usepackage{multicol}
\usepackage{enumitem}
\usepackage{color}

\newcommand\sersic{S\'{e}rsic }
\newcommand\sersics{S\'{e}rsics }
\newcommand\Sersic{S\'{e}rsic}
\newcommand\Chi{$\chi_{\nu}^2$ }
\newcommand\second{^{\prime\prime}}
\newcommand\minute{^{\prime}}

\title[Bulge and Disc Colours of Early-Type Galaxies in the Coma Cluster]
{Dissecting the Red Sequence: The Bulge and Disc Colours of Early-Type Galaxies in the Coma Cluster}
\author[J.T.C.G. Head et al.]
{
Jacob T.C.G. Head$^1$\thanks{Email: j.t.c.head@durham.ac.uk}, John R. Lucey$^1$, Michael J. Hudson$^{2,3}$, Russell J. Smith$^1$\\
$^{1}$Department of Physics, University of Durham, South Road, Durham, DH1 3LE, UK\\
$^{2}$Department of Physics and Astronomy, University of Waterloo, Waterloo, Ontario, N2L 3G1, Canada\\
$^{3}$Perimeter Institute for Theoretical Physics, 31 Caroline St. N., Waterloo, ON, N2L 2Y5, Canada\\
}
\begin{document}

\date{Accepted 2014 February 14. Received 2014 February 11; in original form 2013 December 20}
\pagerange{\pageref{firstpage}--\pageref{lastpage}} \pubyear{2014}
\maketitle{}
\label{firstpage}

\begin{abstract}
We explore the internal structure of red sequence galaxies in the Coma cluster across a wide range of luminosities ($-17$\,$>$\,$M_g$\,$>$\,$-22$) and cluster-centric radii ($0$\,$<$\,$r_{\rm{cluster}}$\,$<$\,1.3 $r_{200}$). We present the 2D bulge-disc decomposition of galaxies in deep Canada-France-Hawaii Telescope $u,g,i$ imaging using GALFIT. Rigorous filtering is applied to identify an analysis sample of 200 galaxies which are well described by an `archetypal' S0 structure (central bulge + outer disc). We consider internal bulge and/or disc colour gradients by allowing component sizes to vary between bands. Gradients are required for $30\%$ of analysis sample galaxies. Bulge half-light radii are found to be uncorrelated with galaxy luminosity ($R_e \sim 1$ kpc, $n\sim2$) for all but the brightest galaxies ($M_g<-20.5$). The S0 discs are brighter (at fixed size, or smaller at fixed luminosity) than those of star-forming spirals. A similar colour-magnitude relation is found for both bulges and discs. The global red sequence for S0s in Coma hence results from a combination of both component trends. We measure an average bulge $-$ disc colour difference of $0.09\pm0.01$ mag in $g-i$, and $0.16\pm0.01$ mag in $u-g$. Using simple stellar population models, bulges are either $\sim2$-$3\times$ older, or $\sim2\times$ more metal-rich than discs. The trend towards bluer global S0 colours observed further from Coma's core is driven by a significant correlation in disc colour with cluster-centric radius. An equivalent trend is detected in bulge colours at a marginal significance level. Our results therefore favour environment-mediated mechanisms of disc fading as the dominant factor in S0 formation.  

\end{abstract}

\begin{keywords}
galaxies: clusters: Abell 1656, galaxies: elliptical and lenticular, cD; galaxies: evolution; galaxies: formation; galaxies: structure
\end{keywords}

\section{Introduction}\label{intro}
Within galaxy clusters, strong radial trends are observed in galaxy morphology (the `morphology-density' relation, \citealp{Dressler1980,Dressler1997}) and star formation rate (the `colour-density' relation, \citealp{Balogh2000a,Lewis2002,Balogh2004,Hogg2004}). This builds a picture in which a galaxy is driven from late (e.g. spiral) to early (e.g. E/S0) morphology as it is accreted onto a cluster due to environment-mediated processes. The current position of a galaxy within a cluster is correlated with its infall time \citep{Gao2004,Smith2012,DeLucia2012,Taranu2012} and velocity \citep{Oman2013}.
Therefore, variations of galaxy structure and colour with cluster-centric radius offer insight into the physical mechanisms which acted during cluster infall, precipitating morphological evolution. 

Structural and kinematical similarities between S0s and spiral galaxies \citep{vdBergh1976,Aragon2006,Bekki2011} imply a evolutionary scenario in which spiral-like progenitors are transformed into early-type galaxies (ETGs). Such transformations would require the erasure of prominent spiral pattern, truncation of the galaxy's star formation, and enhancement of the galaxy's bulge-to-total ratio (B/T $=f_B/(f_B+f_D)$; \citealp{Dressler1980}). Devoid of fuel for continued star formation, the galaxy would then fade and redden as its stellar populations age, moving from the `blue cloud' to the well-known `red sequence'. The resulting `red and dead' descendant galaxy would be bulge-dominated, with a smooth, corotating disc. The environment-mediated mechanisms proposed to drive this morphological transformation can be broadly categorised as either `bulge-enhancing' or `disc-fading'. 

In a `bulge enhancement' scenario, the galaxy's cold gas is rapidly consumed in strong bursts of centrally-concentrated star formation. Additionally, driving the gas towards the galaxy's centre may feed a supermassive black hole, quenching star formation through active galactic nucleus (AGN) feedback \citep{Silk1998,Schawinski2006,Schawinski2007,Brodwin2013}. Mergers (in groups or the field) and galaxy harassment (in clusters, \citealp{Moore1996}) are the most widely studied mechanisms of this type, both of which show the potential to form S0s (\citealp{Bekki1998,Governato2009,Moore1998,Moore1999}). Stellar discs of galaxies are typically disrupted by these processes, however, requiring specific environmental conditions for disc survival \citep{Hopkins2009} or a long period of disc regrowth \citep{Kannappan2009}.
 
Alternatively, a `disc fading' scenario involves either the direct stripping of cold gas from the galaxy's disc (e.g. due to ram pressure \citealp{Gunn1972,Quilis2000}), or removal of its hot halo gas reservoir over a long period of strangulation \citep{Larson1980,Balogh2000a}. These mechanisms act preferentially on gas, causing little or no disruption to the galaxy's stellar disc, but act on different timescales. Additionally, disc fading mechanisms do not act in low-density environments, suggesting that transformation would have had to occur since accretion into a group environment.

Recent work studying galaxy kinematics has indicated that most ETGs are rapidly rotating. In this paradigm, elliptical and lenticular galaxies are part of a continuous sequence of rotating, quiescent galaxies with specific angular momentum increasing in magnitude from E to S0a-c \citep{Emsellem2011,Cappellari2011}. Hence, the S0 morphology encompasses multiple distinct classes of galaxy, differing in luminosity and evolutionary pathway \citep{vdBergh1990,vdBergh2009a}. Conversely, the distinction between S0s and the most elongated (E7) elliptical galaxies is purely due to observational bias \citep{vdBergh2009b}. If so, with sufficient signal-to-noise (S/N), discs should be detectable in most elliptical galaxies. Such multicomponent structures for ETGs are consistent with the predictions of photometric modelling (e.g. \citealp{Rix1990}).

Measurements of broad-band colours can be used to probe a galaxy's star formation history. Using stellar population synthesis models, optical colours can constrain the ages and metallicities of the observed stellar populations, albeit with degeneracies due to the common bias towards redder colours caused by both properties. Many photometric studies of galaxy evolution only investigated {\emph{ global}} colours (e.g. \citealp{Gavazzi2010}). While straightforward, this approach glosses over the rich variation of stellar population properties within and between component structures. Alternatively, bulge-disc decomposition of imaging data (e.g. \citealp{Hudson2010}; hereafter H10, \citealp{Simard2011,Lackner2012}) separates the photometric contributions of the galaxy bulge and disc, and thus enables stellar population properties of these structures to be measured separately. An age difference between the bulge and disc would constrain the transformation mechanisms that drove the galaxy's evolution in the past. 

The well-studied Coma cluster (Abell 1656) possesses one of the richest ETG populations in the local universe. As such, Coma is an ideal laboratory for studying the properties and evolution of ETGs (e.g. \citealp{Lucey1991,BowLucEll,Jorgensen1999}). In addition, Coma encompasses a wide range of local environment conditions, allowing in-depth investigation of radial trends of environment-mediated processes \citep{Gavazzi1989,Guzman1992,Carter2008,Gavazzi2010,Smith2012,Cappellari2013,Rawle2013,Lansbury2014}.

In this work, we investigate the bulges and discs of red sequence galaxies in the Coma cluster within an absolute magnitude range $-22<M_g<-17$. This yields a rich initial catalogue of ETGs for analysis, over a wide range of local environment densities. 

Rather than forcing a bulge + disc morphology on all galaxies regardless of model suitability, we focus exclusively on a sample of galaxies with S0-like structural morphologies (i.e. central bulge + outer disc; described as ``Classic"/Type 1 in \citet{Allen2006}, hereafter `archetypal'). The cherry-picked nature of this sample introduces selection biases as only symmetric galaxies with idealised morphologies are included. However, our aim here is to examine galaxies expected to be well-fit by a bulge + disc model, rather than address a complete sample.

Using high quality Canada-France Hawaii Telescope (CFHT) optical imaging, we perform bulge-disc decompositions with a flexible ($n$ free) \sersic + exponential model using GALFIT \citep{GALFIT}. As this data is significantly deeper than the Sloan Digital Sky Survey (SDSS), with better resolution, our decomposition results achieve a greater level of precision than can be achieved using SDSS imaging. As such, we are able to probe the poorly-investigated faint end of Coma's red sequence with far greater reliability. 

We discuss multi-band fitting (i.e. fitting using constraints derived from fits to another photometric band) based on the results of two alternative methods, differing in their interpretation of galaxy colour gradients: i) where gradients are purely due to a difference in colour between bulges and discs. ii) where gradients are a combination of colour separation and internal component gradients. This builds on earlier multi-band studies (e.g. \citealp{Simard2002,Gadotti2009}), providing the first detailed investigation of internal component colour gradients in the context of 2-component \sersic fits in 2D (although multiwavelength 1-component \sersic fitting is discussed at length in \citealp{Haussler2013}).

We investigate the origins and variation of colour-magnitude trends in ETGs in Coma as a means of probing the mechanisms that influenced their evolution history. We address three main questions: do the bulges and discs of ETGs follow a common red sequence slope?; how separated (in colour) are the stellar populations of these components?; in what way do the observed colour distributions vary during cluster infall? 

The outline of the paper is as follows: An overview of the CFHT imaging, and description of the initial galaxy sample is presented in $\S2$. Details of the bulge-disc decomposition carried out using GALFIT are presented in  $\S3$, including discussion of sample filtering. In $\S4$, we present the structural, colour-magnitude (including fitting performed both with and without internal gradients), and colour-radius results. We present a discussion of the reported results in $\S5$, summarise our findings in $\S6$. A detailed catalogue of fitting measurements are presented in Tables \ref{result_1} and \ref{result_2} of Appendix \ref{Cat}.

Throughout this paper, we use the WMAP7 cosmology: $H_0=70.4\rm{km}\rm{s}^{-1}\rm{Mpc}^{-1}$ (i.e. $h_{70} = 1.01$),$\Omega_m=0.272$ and $\Omega_{\Lambda}=0.728$ \citep{WMAP7}. Using $z_{\rm{CMB}}\rm{(Coma)}=0.024$, the luminosity distance for the Coma cluster is 104.1 Mpc, and the distance modulus, $m - M = 35.09$. At this distance, $1\minute$ corresponds to 28.9 kpc. Taking a value for velocity dispersion of $\sigma_{{\rm Coma}} = 1008$ km s$^{-1}$ \citep{Struble1999} and virial mass, $M_{200} = 5.1\times10^{14} h^{-1}_{70}$\Msol \citep{Gavazzi2009}, the virial radius, $r_{200}$, for Coma is 2.2 Mpc ($\sim75\minute$).

\section{Data and Initial Sample}\label{samp}
Optical imaging covering a total of 9 deg$^2$ of the Coma cluster in the $u$-, $g$-, and $i$-bands was acquired using the MegaCam instrument on the 3.6 m CFHT during March - June 2008 (Run ID 2008AC24, PI: M. Hudson). Total (coadded) exposure times of 300 s were obtained for the $g$- and $i$-bands, and 1360 s for the $u$-band. Compared to SDSS (2.5 m telescope, 53 s exposures), these observations were $\sim12\times$ deeper in the $g$- and $i$-bands, and $\sim50\times$ deeper in the $u$-band (from $D^2t_{\rm exp}$). We note that the MegaCam $u$ filter is shifted slightly redward relative to SDSS $u$. This difference is accounted for by calibrating the $u$-band zero points with SDSS aperture photometry (see Section \ref{prep}). The MegaCam frames were sky-subtracted during pre-processing using a 64 pixel mesh. A point spread function (psf) full-width half-maximum (fwhm) of between $0.65\second$ and $0.84\second$ was typical. The pixel scale was $\sim0.186$ arcseconds/pixel.

The initial sample for analysis was selected from SDSS (DR9) catalogue galaxies in the 3 deg $\times$ 3 deg ($\equiv5.2$ Mpc $\times 5.2$ Mpc, 2 $r_{200} \times$ 2 $r_{200}$) area covered by the MegaCam observations. A limit of $M_g<-17.1$ was applied to ensure sufficient signal-to-noise (S/N) for reliable measurement of galaxy bulge and disc structures. Likewise, a colour limit of $(g-r)>0.5$ was used to isolate red sequence galaxies ($N=69$ removed). These targets were limited to the redshift range $0.015 < z < 0.032$ (heliocentric $v_{\rm{Coma}} \pm2.5\sigma_{\rm{1D}}$) to ensure that only cluster members were included. These selection criteria yield a sample of $\sim600$ red-sequence galaxies.

\section{Analysis}
To measure the structural and photometric parameters of galaxy bulges and discs, galaxy decomposition has been carried out using GALFIT (version 3.0.4), a 2D fitting routine \citep{GALFIT}. Given a user-specified model (of arbitrary complexity), GALFIT varies parameters based on a non-linear chi-squared minimisation algorithm until no significant reduction in chi-squared ($\chi_{\nu}^2$) is found. The parameter values of this best-fit 2-component model are used to estimate the underlying structure and photometry of the target galaxy. 

Throughout this paper, we do not include a formal correction for the presence of dust internal to the observed galaxies. Dust effects will contribute to the scatter in structural and photometric distributions presented below. This effect is pronounced in highly inclined galaxies, but will also influence face-on galaxies \citep{Driver2007}. By filtering our analysis sample to remove galaxies with strong dust lanes or asymmetries (see Section \ref{decomp}), we have minimised the effects of dust on our conclusions. The detection and correction of remaining dust-contaminated galaxies (via Spitzer data) will be the subject of a future study.

\subsection{Initial Fitting Preparation}\label{prep}
For GALFIT's primary data input, $\sim100\second\times 100\second$ (536 $\times$ 536 pixel) thumbnail images were extracted from the coadded MegaCam image frames, centred on each target galaxy. The local background sky was measured by fitting isophotal ellipses, tracing the surface brightness down to a level 5\% above the sky. As no significant variation ($<1\%$) is detected between sky measurement annuli, the sky value is assumed to remain constant for each thumbnail.

Galaxy thumbnails are masked during fitting to prevent the sky from dominating the value of $\chi_{\nu}^2$. SExtractor \citep{SEX} was used to both identify pixels containing flux from the target galaxy (at a $0.5\sigma_{\rm{sky}}$ threshold above the sky), and to detect and mask other sources or defects in the thumbnail ($>\sigma_{\rm{sky}}$). The elliptical area containing the galaxy (hereafter `target ellipse') was extended by $5\second$ along the semi-major axis (with $b/a$ fixed) to ensure that no target flux was omitted. All pixels outside of the target ellipse were masked, except for $\sim1\second$ wide strips along its major and minor axes. This unmasked cross area was included  to avoid discontinuities at the mask boundary. 

Noise maps were produced to ensure correct weighting of each pixel during $\chi^2$-minimisation. The underlying noise distribution, $\sigma(x,y)$,  was estimated using the method given in \cite{GALFIT}. However, the mean sky flux and the rms sky variance were estimated independently to avoid contamination by low surface brightness sources. 

Stars in the MegaCam frames were used to characterise the psf of the data images. The closest five stars (of sufficient S/N) were averaged using KAPPA and MAKEMOS (from the Starlink package, \citealp{Starlink}) to produce a master psf image for each band. Stars were selected no further than $5\minute$ from the target galaxy, and were vetted to exclude unsuitable (e.g. saturated, binary) images.

The magnitude zero points were calibrated on a galaxy-by-galaxy basis by comparing SDSS photometry ($7.43\second$ aperture) with the magnitude measured in an equivalent radius aperture in the MegaCam images. This correction allows direct comparison of this work's results with studies utilising the Sloan $ugriz$ magnitude system \citep{Fukugita1996,Smith2002}. To first order, this also accounts for the differences between MegaCam and SDSS $u$-band filters. The zero point uncertainty in all bands was typically $\lesssim0.01$ mag.

Absolute rest-frame magnitudes were calculated by subtracting the distance modulus ($m-M =35.09$), and applying galactic dust extinction- and $k$-corrections. Using the maps of \cite{Schlafly2011}, galactic extinction corrections of 0.034, 0.026, and 0.014 mag were applied to the $u$-, $g$-, and $i$-bands respectively. The $k$-correction was calculated using SDSS spectroscopic redshifts and ``the $k$-correction calculator" \citep{KCorr,Kcorr2}. This correction was typically 0.12, 0.05, and 0.01 mag in the $u$-, $g$-, and $i$-bands.

\subsection{Image Decomposition Procedure}\label{decomp}
We fit a \sersic (\citealp{sersic}, hereafter `bulge') + exponential ($n=1$, hereafter `disc') model to all galaxies in the Coma sample. A preliminary one-component (\Sersic) model fit is used to provide starting values for the 2-component fitting parameters. This iterative build-up of model complexity reduces sensitivity of the measured parameters to the choice of input values. GALFIT's search through parameter space was extended to improve the reliability with which the true global minimum in $\chi_{\nu}^2$ space was found (for further details see Appendix \ref{AGONII}). 

The aim of this study is to characterise the photometric and structural properties of the bulges and discs of ETGs. The Coma sample (as defined in Section \ref{samp}), however, contains many galaxies that are poorly described by an archetypal bulge + disc structural morphology. Such galaxies are selectively removed from further analysis by a sequence of logical filters to ensure the applicability of a `bulge + disc' interpretation. 

First, asymmetric, contaminated, or poorly-fit galaxies (i.e. high $\chi^2$) were removed from the sample. This includes cases where galaxy crowding (prominent towards the cluster core) would require additional deblending steps to accurately model. These fits (hereafter `unstable' sample) do not yield reliable model parameters, and therefore cannot be considered accurate representations of the underlying galaxy structures.

Secondly, galaxies best fit by a single-component model were excluded. These \Sersic-only galaxies were identified where the addition of the exponential disc component provided no significant improvement to the model fit. The comparative goodness-of-fit of the models was assessed using the Bayesian Information Criterion (BIC, \citealp{BIC}). 

Thirdly, model fits which yielded extreme values of B/T were considered indistinguishable from pure \sersic systems due to high fitting parameter uncertainty ($>10\%$) in components contributing $\leq10\%$ of the total galaxy luminosity. The sample of 1-component galaxies (as identified by either BIC or B/T) is referred to as the `\Sersic' sample hereafter.

\begin{figure}
\begin{center}
	\includegraphics[width=\linewidth,clip=true]{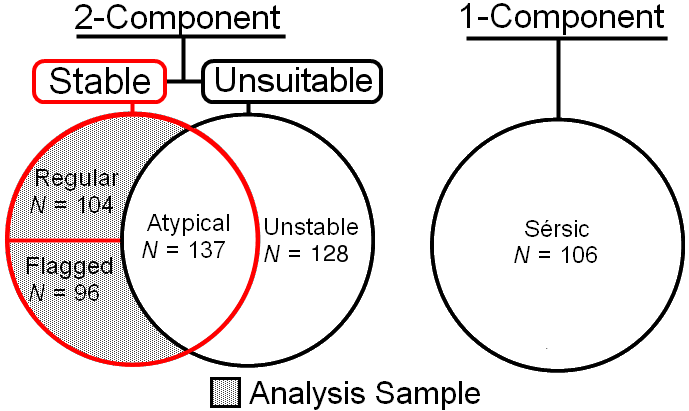}
\end{center}
\caption{Venn diagram describing the main filtering samples and sub-samples (see Table \ref{filt_tab}) for galaxies best fit by \sersic + disc or \Sersic-only models.}
	\label{Venn_samp}
\end{figure}

\begin{table}
\begin{tabular}{clc}
\hline
\hline
Sample & Subsample & $N$\\
\hline
\hline
Initial & All Galaxies Fitted & 571 (100\%) \\
\hline
Analysis & Regular Fit & 104 (18.2\%)\\
Sample&  Flagged (High $\chi^2$) & 52 (9.1\%)\\
$N=200$& Flagged (Bar Component) & 33 (5.7\%)\\ 
(35.0\%) & Flagged (Swapped Profile) & 11 (1.9\%)\\
\hline
 & Atypical Profile & 137 (24.0\%) \\
Unsuitable &  Asymm./Spiral/Dust Struct.& 56 (9.8\%)\\ 
B+D Fits & Contaminated/Defective & 23 (4.0\%)\\
 $N=265$& Bar/Twist/Edge-on & 17 (3.0\%)\\
(46.4\%) & Ring/Peanut Structure & 16 (2.8\%)\\ 
 & Blue Core Col. Profile & 16 (2.8\%)\\
\hline
\sersic Sample&  \sersic (BIC) & 88 (15.4\%) \\
$N=106$&  \sersic (B/T) & 12 (2.1\%)\\
(18.6\%)& \sersic (Boxy) & 6 (1.1\%)\\
\hline
\hline
\end{tabular}
\caption{The samples (and subsamples) of galaxies resulting from applying the logical filter (see Section \ref{decomp}) to the fitting results. The number of galaxies occupying each sample, $N$, is given along with the percentage fraction of the initial data sample.}
\label{filt_tab}
\end{table}

The remaining galaxies are well-fit by a 2-component bulge + disc model (i.e. stable 2-component fit). Surface brightness profile types (as formalised in \citealp{Allen2006}) are used to separate archetypal (i.e. inner bulge, outer disc) galaxies from those with atypical morphologies. Additionally, a fraction of archetypal galaxies have been highlighted as stable, but lower-quality fits (hereafter `flagged' sample). This includes fits which suggested the presence of additional structural components (typically central bars), which would distort interpretation of bulge and disc component colours. A few ($N=12$) galaxies are also flagged where an archetypal morphology could only be achieved by manually swapping the \sersic and exponential structural parameters. All non-flagged, archetypal 2-component galaxies are described as the `regular' sample hereafter.  

The results of this sample filtering procedure are summarised in Figure \ref{Venn_samp} and Table \ref{filt_tab}, and described in detail in Section \ref{filt_res}. A full description of the logical filter and associated statistical tests is given in Appendix \ref{filter}.

\subsection{Multi-band Fitting}\label{mbfit}
ETGs (particularly ellipticals) are well-known to possess internal radial gradients in colour (e.g. \citealp{Vader1988, Franx1990}). These colour gradients result from variation of stellar population properties; primarily metallicity. Gradients of $-0.1$ mag dex$^{-1}$ in $B-R$ ($\approx g-i$) and $-0.2$ mag dex$^{-1}$ in $U-R$ are typical \citep{Wu2005}.

For a 2-component galaxy model, galaxy colour gradients can be interpreted as the transition between inner- and outer-dominating structures with differing component colours. The detection of bluer outer discs than inner bulges thus reflects the typical negative (i.e. bluer with increasing galaxy radius) galaxy colour gradients. Alternatively (or in addition), the structural components may possess internal colour gradients due to variation of stellar population properties with galaxy radius.

\begin{figure}
\begin{center}
	\includegraphics[width=\linewidth,clip=true]{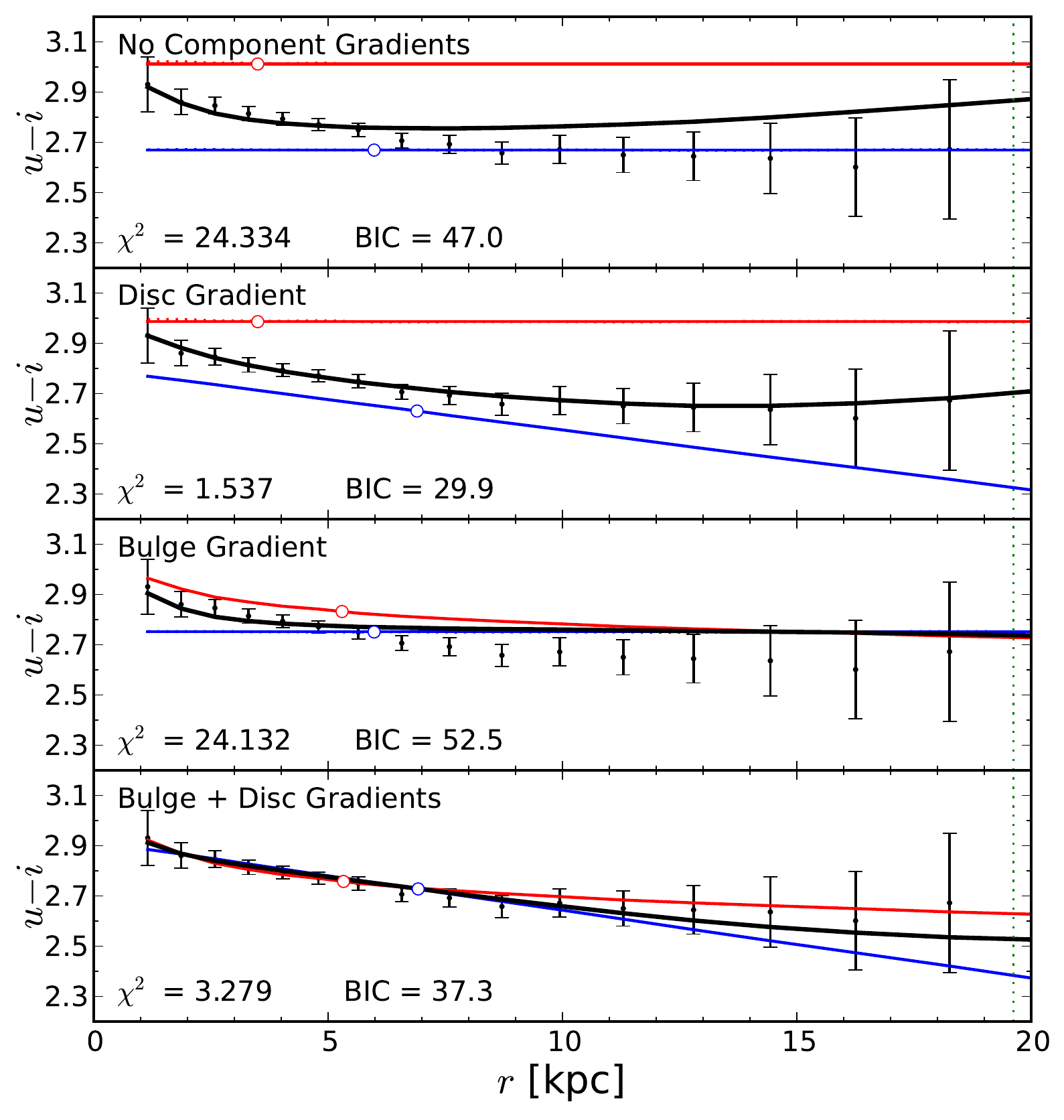}
\end{center}
\caption{$u-i$ colour profiles for a typical galaxy (SDSS DR8 ID: 1237665440979026019) as measured in elliptical annuli. Black error bars indicate the colour measured from the galaxy thumbnail, while the derived model bulge, disc, and total colour profiles are plotted as red, blue and black solid lines. Unfilled red and blue points indicate the global component colours. From top to bottom, the different panels show colour profile models with no component gradients, disc gradients only, bulge gradients only, and both bulge and disc gradients. For each profile, the (non-reduced) 1D chi-squared, and corresponding BIC statistic is included. Vertical green dotted lines indicate the edges of the object ellipses used during masking.} 
	\label{grad0}
\end{figure}

\begin{figure}
\begin{center}
	\includegraphics[width=\linewidth,clip=true]{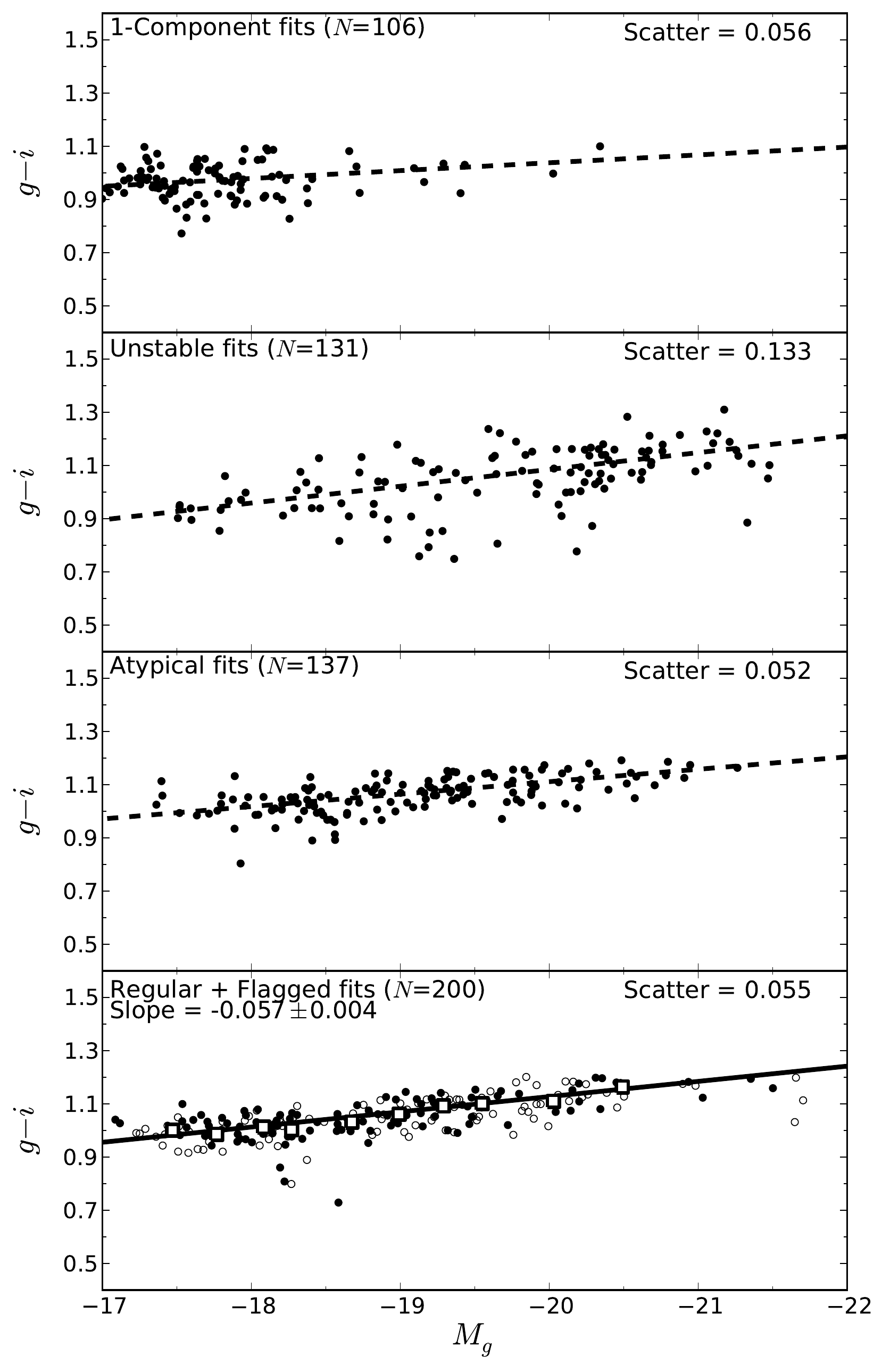}
\end{center}
\caption{The measured (GALFIT) model $g-i$ colours plotted against their absolute (bulge + disc) $g$-band magnitudes. {\bf Top Panel:} Galaxies best fit with a \Sersic-only model. {\bf Upper-middle Panel:} Galaxies with no stable 1- or 2-component fit. {\bf Lower-middle Panel:} 2-component galaxies which do not possess an archetypal central bulge + outer disc morphology. {\bf Bottom Panel:} Regular and flagged (unfilled points) fits. The large square points indicate median colour values calculated in magnitude bins, to which a linear trend (solid line) is fitted. The slope uncertainty is calculated from bootstrap fitting the unbinned data. This best fit line is plotted for comparison as a dashed line in the other three panels.}
	\label{sample_CM}
\end{figure}

\subsubsection{Fixed Multi-band Fitting}
A straightforward method of fitting a galaxy across multiple photometric bands would be to fit each band entirely independently (e.g. \citealp{Gadotti2009}). In practice however, model degeneracies often yield inconsistent model structures if the fitting is not appropriately constrained. In order to ensure that the photometry is being measured reliably, consistent structures must be fit to the galaxy in each band.

The most basic method of implementing multi-band fitting is to adopt a common structure for the models of all fitted bands. This `fixed multi-band' approach is carried out by determining the structural parameters (\sersic effective radius, $R_e$, \sersic index, $n$, exponential scale length, $Rs$, position angle, PA, component axis ratio, $b/a$) from one freely-fit band, and fixing their values for fits to other bands. These `dependent' fits allow only the component magnitudes to vary, with both components sharing a common centre. Here, the $g$-band is selected for `free' fitting (i.e. all structural parameters unfixed).

The limitation of fixed multi-band fitting is that no allowance is made for variation in surface brightness profile shape from band to band, and thus internal colour gradients in the bulge or disc are precluded. Where this approach fails to adequately describe a galaxy's colour information, fitting can be improved by introducing band-dependent variation of structural parameters. 

\subsubsection{Component Gradients}\label{colgradintro}
If fixed multi-band fitting is not sufficient to reproduce the overall colour gradient of a galaxy, internal component colour gradients may be necessary. In the following section, we describe the use of bulge and/or disc colour gradients to more accurately model the galaxy across the $u$-, $g$-, and $i$-bands.

The goodness-of-fit of galaxy colour gradients resulting from multi-band fitting was assessed using observed 1D radial colour profiles. Free fitting was carried out in the $g$-band to determine the model structure. Colour profiles are measured numerically from the thumbnails and model images of the dependent $u$- and $i$-bands. For robust measurement of the colour profile, identical annuli and a consistent resolution must be used across all photometric bands. Resolution-matching was carried out by smoothing each band's psf image such that the modified full-width at half maximum was equal to the largest (unmodified) fwhm of the three band's psfs. The same smoothing was then applied to the data and model images in each band.

Colour gradients are added to the model components by allowing systematic band-to-band variations in $R_e$ for bulge gradients, or $R_s$ for disc gradients. As $R_e$ and $n$ are covariant, $n$ is held fixed where bulge gradients are included to avoid this degeneracy. Increasing the number of free parameters in the `dependent' bands reduces the reliability of the fit by introducing additional model degeneracy. To minimise this problem, a 1D BIC is calculated to select the simplest model (i.e. fewest free parameters) that sufficiently represents the underlying colour profile (see Appendix \ref{filter} for details). 

Example colour profiles (with and without component gradients) for a typical galaxy are presented in Figure \ref{grad0}. In this case, the measured galaxy colour information in the disc-dominated region of the galaxy ($r > 6$ kpc) is only reproduced when a disc gradient is permitted. The 1D BIC shows no significant further improvement if both bulge and disc gradients are included. Alternatively, this outer region of the galaxy colour profile could be reproduced if the disc of the model with no component gradients (Figure \ref{grad0}, top panel) was shifted to a bluer colour. However, this shift would worsen the fit to the more precisely known inner region of the profile ($2 < r < 6$ kpc), yielding a poorer fit overall. 

A small number of galaxies ($N=16$) exhibited colour profiles which become bluer towards the galaxy core, but possessed a negative colour gradient (bluer with increasing $r$) overall. Such systems are more complicated than can be reproduced by a simple two-component model (with or without component gradients), and were removed from the analysis sample for this reason. Adopting colour gradients in at least one component is necessary for $\sim1/3$ of the sample (see Section \ref{grads}). The intrinsic scatter of the measured colours was increased due to the degeneracies introduced as more model parameters are allowed to vary.

\section{Results}
\subsection{Initial Fitting Overview}\label{res_over}
\subsubsection{Sample Filtering}\label{filt_res}
First, we report the global characteristics of all galaxies in the initial Coma sample, and describe the selection of the final analysis sample. Except where specifically noted (i.e. during discussion of internal component gradients), galaxy structural parameters described in all following sections are measured from the freely-fit (i.e. no fixed parameters) $g$-band images.

The filtered sample sizes are presented in detail in Table \ref{filt_tab}. We find a stable fitting solution (either a \sersic-only or \sersic + exponential model) for 443 of the 571 galaxies in the initial Coma sample. Evidence for a significant second structural component is found in 76\% ($N=337$) of these galaxies. `Archetypal' (central) bulge + (outer) disc structures are found in 59\% of 2-component galaxies ($N=200$). The remaining 41\% represent `atypical' 2-component structural morphologies (e.g. centre-dominating bulges which re-dominate the outer regions, dominant or sub-dominant discs at all radii, or inverted disc-bulge structures). Across the entire Coma sample, stable 2-component structures (of any type) are detected in 59\% of galaxies, while 35\% correspond to stable, `archetypal' galaxies.

A total of 200 galaxies are found to be well-fit by an archetypal bulge + disc morphology, including 96 galaxies (48\%) flagged as lower-quality fits (see Section \ref{decomp}). Of these archetypal galaxies, 60 ($30\%$) require the inclusion of colour gradients in at least one model component (see Section \ref{colgradintro}). Nine ($4.5\%$) require bulge gradients, $40$ ($20\%$) require disc gradients, and $11$ ($5.5\%$) require both bulge and disc gradients. 

\begin{figure*}
\begin{center}
	\includegraphics[width=\linewidth,clip=true]{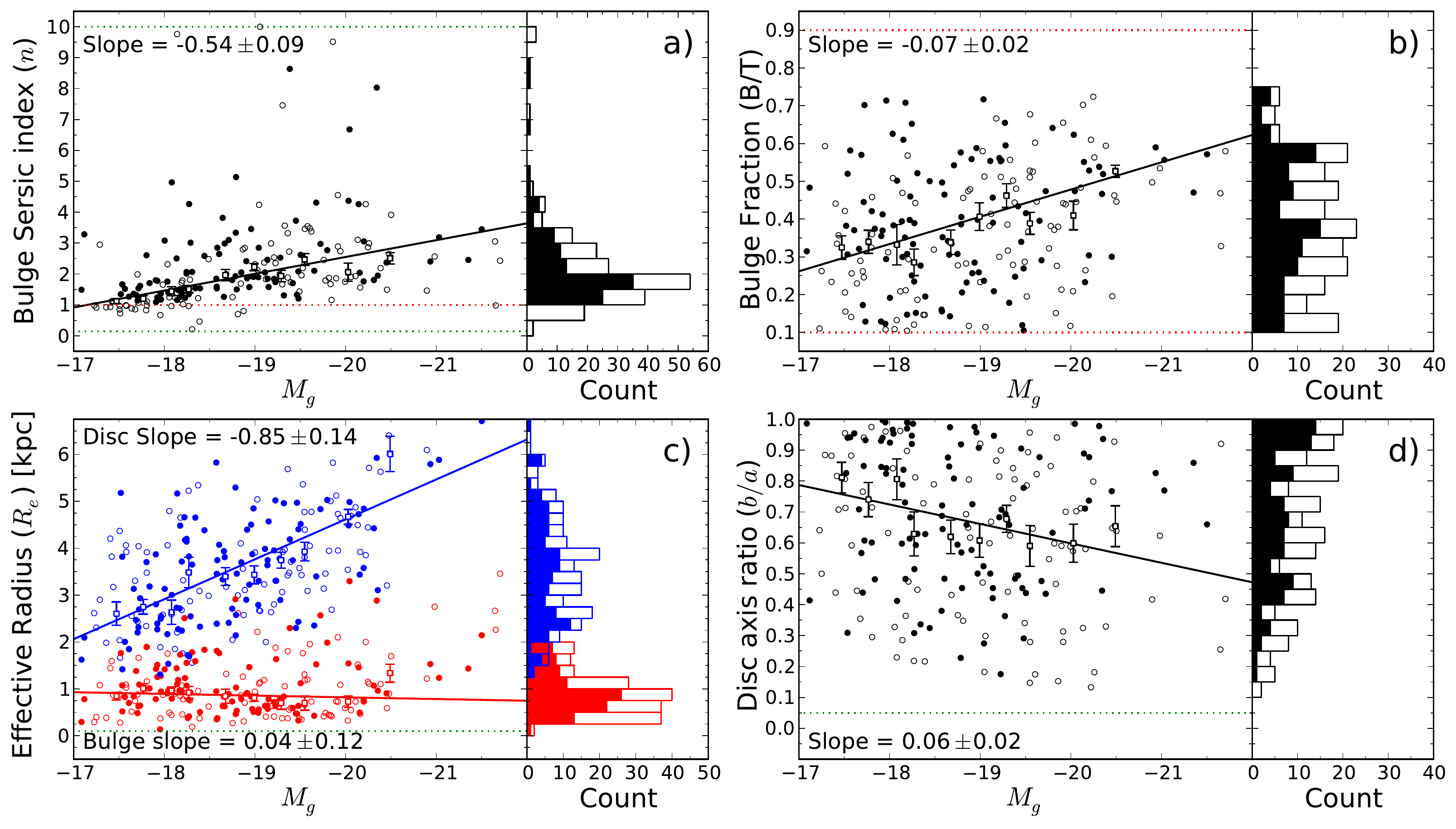}
\end{center}
\caption{The measured structural parameters of (freely-fit) $g$-band (analysis sample) galaxy models plotted against their absolute (bulge + disc) magnitudes. Unfilled points indicate `flagged' fits (see Appendix \ref{filter}). Best-fit linear trends (solid lines) are fit to median parameter values calculated in magnitude bins (large square points), weighted by the robust standard error in each bin (error bars). The slopes and their bootstrap errors (calculated from the unbinned data) are included in each plot. {\bf Top left:} Bulge \sersic index, $n$. {\bf Top Right:} Bulge-to-total luminosity ratios, B/T (bulge fractions). {\bf Bottom left:} Effective half-light radii of bulges ($R_{e,B}$; red) and discs ($R_{e,D}$; blue). {\bf Bottom right:} Disc axis ratio, $b/a$. Histograms of the structural parameter distributions are presented in the right-hand panel of each plot. The unfilled portion of each histogram bar corresponds to the number of `flagged' fits in that bin. Dashed lines in each plot indicate any constraints on parameters values (green: hard limits imposed on GALFIT; red: sample filtering cuts).}
	\label{g_structure}
\end{figure*}

The ratios of measured $R_e$ in the $u$-:$g$-:$i$-bands in models requiring bulge gradients were 1.14:1.00:0.94 on average. The equivalent average $u$:$g$:$i$ ratios of $R_s$ in models requiring disc gradients were 1.03:1.00:0.98. \cite{Kelvin2012} presented wavelength-dependent variation of structural parameters for single \sersic fits of disc- and spheroid-dominated galaxies. Calculating $R_e$ and $R_s$ using median passband wavelengths yields $u$:$g$:$i$ ratios of 1.08:1.00:0.87 and 1.05:1.00:0.92 respectively. Thus, the internal component gradients included in this work are similar to previous measurement of size variations for bulge- and disc-dominated galaxies. The component size variations are systematically smaller than global size variations measured by \citeauthor{Kelvin2012} (with the exception of the $u$-band bulge $R_e$). This reflects the shallower colour gradients of bulge and disc components relative to their parent galaxy as a whole.

\subsubsection{Filtered Sample Properties}\label{filt_prop}
The global (bulge + disc) colour-magnitude properties of the `\Sersic', `unstable', `atypical', and `archetypal' filtering samples are presented in Figure \ref{sample_CM}. `Regular' and `flagged' subsample galaxies are plotted in the lower panel as filled and unfilled points (respectively) to highlight the results of lower-quality fits. 

Galaxies in both the `regular' and `flagged' samples follow a common red sequence, with similar scatter in both samples (0.058 mag and 0.054 mag respectively). Atypical galaxies exhibit a consistent red sequence slope and scatter. Thus, while a forced bulge + disc interpretation of atypical galaxies would yield incorrect component photometry, their global properties are consistent with archetypal galaxies. `Unstable' sample galaxies, by contrast, appear bluer on average (e.g. due to the prevalence of spiral morphologies), and possess a far larger scatter in colours. 

Galaxies classified as pure \sersics ($n=1.9\pm0.1$ on average) are mainly found at $M_g >-18.5$. This result partly reflects the prevalence of dE and dS0 galaxies at the faint end of the magnitude range, as detected in Virgo ($M_g > -18$, \citealp{Sandage1985}) and Coma ($M_R > -19.5 \approx M_g > -18.5$, \citealp{Aguerri2005}). Misclassification of faint 2-component galaxies due to the low S/N remains a possibility, however, especially for highly bulge- or disc-dominated galaxies. Nevertheless, the absence of classic (1-component) giant ellipticals supports a multi-component structure paradigm for ellipticals (see e.g. \citealp{Huang2013a}). These additional components are not necessarily classical discs, and may require additional model components to be fully characterised. The scatter of `\Sersic' sample galaxy colours is similar to the `archetypal' sample (0.056 mag vs 0.055 mag). 

In the following analysis, we exclude `unstable' and `atypical' fits (hereafter ``unsuitable" sample), and the pure \sersic sample. The remaining 200 `archetypal' \sersic + exponential galaxies in the `regular' and `flagged' samples are described as the ``analysis sample" hereafter. 

\subsection{Bulge and Disc Model Structures ($g$-band)}
In this section, we describe the distributions of bulge and disc structural parameters for galaxies in the analysis sample. A full catalogue of structural fitting measurements are presented in Table \ref{result_1} of Appendix \ref{Cat}.

Galaxy bulges exhibit the full range of permitted \sersic indicies ($0.15 < n < 10$), although bulges with $n \gtrsim 4$ (i.e. more centrally concentrated) are rare, and the majority fall within the range $n \sim 1 - 3$ (Figure \ref{g_structure}a). The median bulge $n$ is $1.86\pm0.11$. This distribution of \sersic indices is similar to those reported in \cite{Krajnovic2013} for ATLAS3D `fast rotator' galaxies ($\sim75\%$ within $0.5<n<3$). Conversely, GIM2D bulge-disc decompositions of a broader sample of galaxies in \cite{Simard2011} yielded a more even spread of bulge $n$ in the range $0.5<n<8$ (peaking at $n\sim6$). 

The measured bulge fractions of the sample are consistent with a flat distribution in the range $0.1 < \rm{B/T} \leq 0.6$ (median B/T $=0.38\pm0.01$), but the occupancy of bins above B/T$\sim0.6$ drops sharply (Figure \ref{g_structure}b). This lack of bulge-dominated galaxies partially reflects an underlying deficit in the initial (i.e. unfiltered) distribution of B/T, but is exacerbated by sample filtering. If `atypical' fits are {\emph{included}}, the combined bulge fraction distribution becomes consistent with the distribution measured for the initial (i.e. unfiltered) sample. The (B- and R- band) decompositions in H10 for cluster galaxies exhibit a sharp peak for $0.4<$B/T$<0.6$, with greater occupancy of the $0.6<$B/T$<1.0$ bin, and lesser occupancy of the $0.0<$B/T$<0.4$ bin than in the present work. This lower median B/T reflects the higher magnitude limit (i.e. fainter galaxies) in this work relative to H10. B/T distributions of \cite{Krajnovic2013} fast rotators exhibit a peak in the range $0.05<$B/T$<0.40$, but are dominated by the $0.95<$B/T$<1.00$ bin ($\sim36\%$).

The physical sizes of bulges are distributed across a narrow range of effective half-light radii ($0.5 \leq R_{e,B} \lesssim 1.5$ kpc), with a median value of 0.9 kpc (Figure \ref{g_structure}c). 
Galaxy discs are larger (median $R_{e,D}=3.6$ kpc), and spread over a wider range of effective radii ($2.0 \lesssim R_{e,D} \lesssim 6.0$ kpc). Note that the effective half-light radius of an exponential disc is related to the scale length parameter ($R_s$) used during fitting as $R_{e,D} = 1.678R_s$. For comparison \citeauthor{Huang2013a} (\citeyear{Huang2013a}; hereafter H13) performed 3-component \sersic fits to $\sim100$ Carnegie-Irvine Galaxy Survey elliptical galaxies across a range of environments. The bulge \sersic indices measured in our study are distributed similarly to the `inner' (median $n\sim2.0$) and `outer' (median $n\sim1.6$) H13 \sersic structures, with sizes approximately consistent ($<2\sigma$) with the former ($R_{e,B} \sim 0.6$ kpc). Our disc sizes fall between the median value of `middle' ($R_{e,B}\sim2.5$ kpc) and `outer' ($R_{e,B}\sim10.5$ kpc) H13 structures. Thus the bulge components in our study are equivalent to the `inner' H13 structures, while the discs in this work are a combination of the `middle' and `outer' components.

The $R_{e,B}$ and $n$ values measured here for bulges are similar to dwarf ellipticals \citep{Graham2003} and giant S0s and early type spirals \citep{Aguerri2004} in Coma. A more general study of ETGs in the core of Coma \citep{Gutierrez2004} found a broader range of bulge sizes due to the presence of giant ellipticals in the sample. Nevertheless, this E+S0 sample yielded average $R_{e,B}$ and $n$ values consistent with our results. Additionally, as noted in other relevant works (\citealp{Graham2013}, H13), these bulges are comparable in structure with ``red nugget" galaxies (e.g. \citealp{Buitrago2008,Damjanov2009}) observed at $z>1.5$ ($R_{e,B} \sim 1$ kpc, $n\sim2$). This similarity between S0 bulges and high-redshift, compact galaxies may indicate formation of S0 progenitors via rapid disc-growth around pre-existing bulges between $z\approx1.5$ and the epoch of quenching. However, current stellar mass estimates for red nuggets are approximately an order of magnitude larger than the mass of typical Coma galaxy bulges. Hence, a scenario in which discs are assembled around existing red nuggets is disfavoured.

The measured structural parameters correlate with the total luminosity of the galaxy. More luminous galaxies are more bulge-dominated (Figure \ref{g_structure}b), with more centrally concentrated (larger $n$) bulges (Figure \ref{g_structure}a), and larger discs (Figure \ref{g_structure}c). No significant size-luminosity trend is detected within galaxy bulges. Thus, the process(es) responsible for building mass in these galaxies caused significant disc growth, with relatively little change to their bulge sizes. No significant change is noted in the size-luminosity slopes if component sizes are compared to component magnitude instead of total magnitude. By comparison, \cite{Laurikainen2010} found a significant correlation between $R_{e,B}$ and (bulge) $K$-band luminosity for S0s. However, their S0 sample is intrinsically brighter ($M_K$ (bulge) $<-21 \approx M_g$ (total) $< -20$) than the present work. Thus, the size-luminosity trend for S0 bulges is flat for the range of galaxy luminosities studied here ($M_g > -20.5$) with a significant upturn for brighter S0s (see also \citealp{MendezA2008,Hyde2009,Allanson2009}). 

\begin{figure}
\begin{center}
	\includegraphics[width=\linewidth,clip=true]{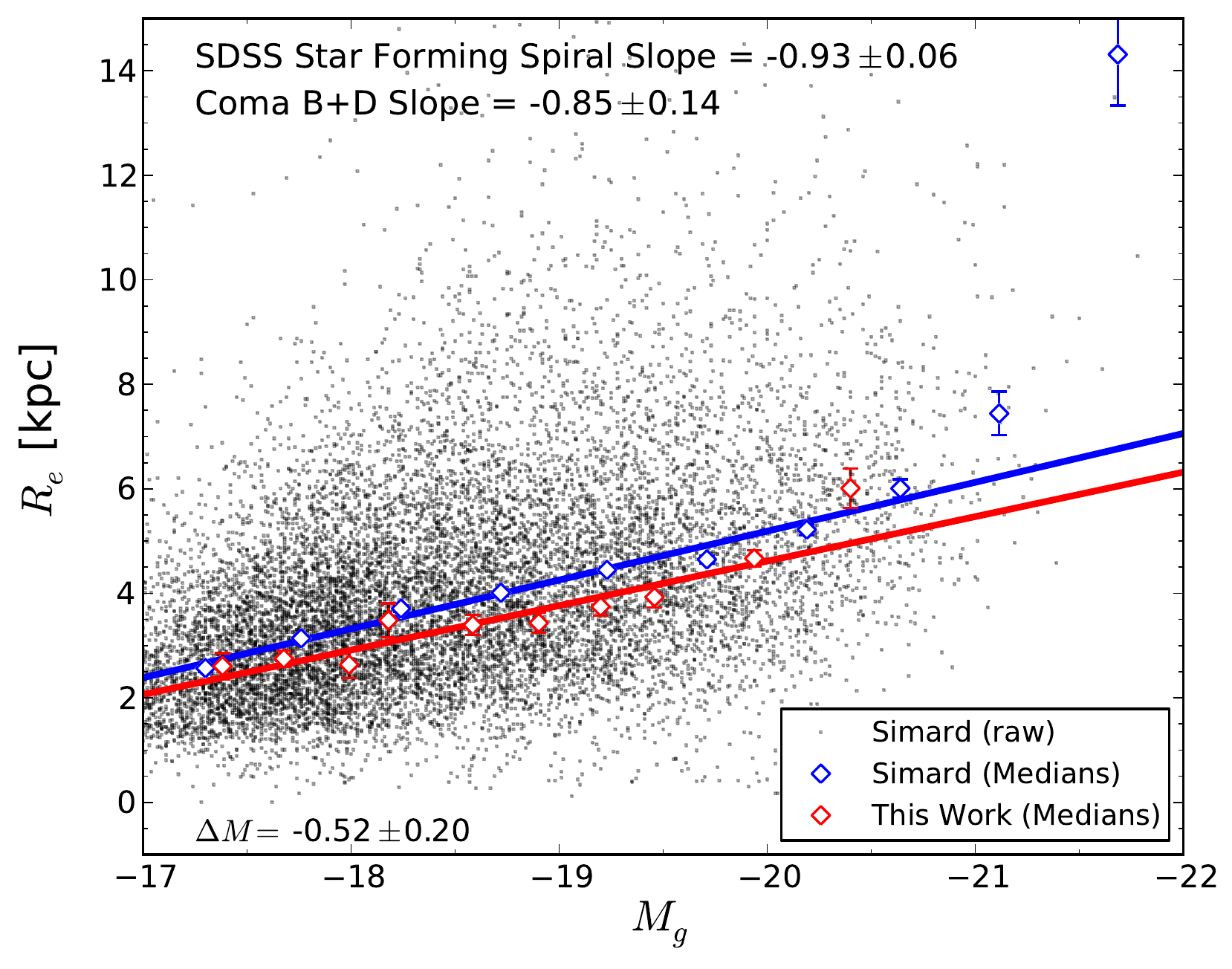}
\end{center}
\caption{
The size-luminosity relation ($R_{e,D}$ [kpc] vs. absolute bulge + disc $g$-band magnitude) for S0s (this work; red) and star forming spirals (black). Spiral galaxies are plotted using bulge-disc decomposition parameters from \protect\cite{Simard2011} for all galaxies at $z < 0.03$ where a debiased probability of spiral morphology (Galaxy Zoo, \citealp{GalaxyZoo}) is greater than 0.5. Star forming galaxies were selected using the MPA-JHU DR7 star formation catalogue \citep{Brinchmann2004} where the specific star formation rate, ${\rm{log}}($\Msol$/{\rm yr}/M_{*}) > -10.5$. Median disc size values are measured in bins of absolute magnitude (0.5 mag width), to which best fit lines are fit. The slope uncertainties are bootstrap errors calculated from the unbinned data. This robust fitting is carried out over the same magnitude range for both datasets.
}
	\label{siz_lum}
\end{figure}

The disc size-luminosity slope ($R_{e,D}$ vs $M_g$) measured in the present work is $-0.85\pm0.14$ kpc mag$^{-1}$, consistent with the equivalent observed slope for star-forming spiral galaxies measured for the same magnitude range ($-0.93\pm0.06$ kpc mag$^{-1}$; see Figure \ref{siz_lum}). This spiral sample was selected from \cite{Simard2011} using Galaxy Zoo morphology \citep{GalaxyZoo}, and MPA-JHU DR7 star formation rates. A marginal offset in total magnitude is noted between the best fit trends of S0s and spirals ($-0.52\pm0.20$ mag; i.e. for a fixed disc size, `archetypal' S0 galaxies are 0.52 mag brighter than star forming spirals). If the size-luminosity trend is instead measured in terms of disc luminosity, the slopes for `S0s' and spirals do not change significantly, but the luminosity offset is reduced to $-0.27\pm0.12$ mag. 

The slope of the B/T-luminosity trend measured here ($-0.07\pm0.02$) is consistent with the R-band slope in H10 ($-0.069\pm0.020$) despite differing B/T distributions. This indicates a significant trend of increasing bulge dominance for brighter red sequence galaxies. 

The discs' apparent axis ratios are approximately equivalent to the cosine of their inclinations to the line-of-sight. The measured axis ratios are consistent with the expected flat distribution of inclinations down to $b/a = 0.3$ (Figure \ref{g_structure}d). Significantly fewer discs with smaller axis ratios (more inclined) are detected, particularly for well-fit galaxies. This deficit of highly inclined galaxies is apparent even if the initial sample is not filtered, and a similar distribution of \Sersic-only axis ratios is measured. This indicates a systematic bias in the fitting against high ellipticity (i.e. edge-on) structures, and/or reflects the intrinsic thickness of S0 discs. 

\begin{figure}
\begin{center}
	\includegraphics[width=\linewidth,clip=true]{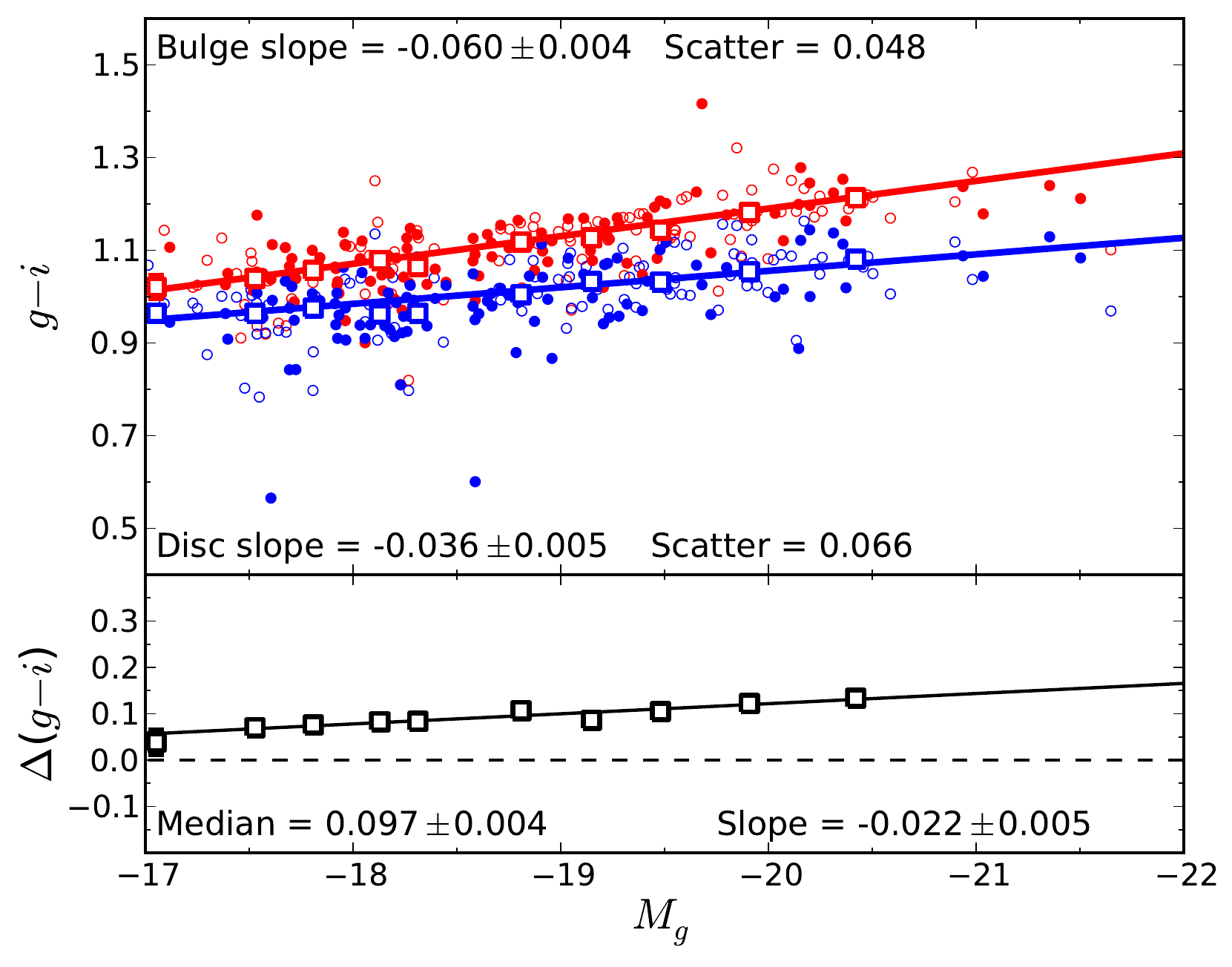}
\end{center}
\caption{Colour-magnitude diagram for fixed multi-band structure. {\bf{Upper panel:}} The bulge (red) and disc (blue) $g-i$ colours plotted against the absolute (bulge + disc) $g$-band magnitudes of each galaxy. {\bf{Lower panel:}} The $g-i$ colour difference (bulge $-$ disc) plotted against absolute $g$-band magnitude. Unfilled points indicate `flagged' fit (containing weak extra components or asymmetries, see Appendix \ref{filter}). Best-fit linear trends (solid lines) are fit to median colour values calculated in magnitude bins (large square points), weighted by the robust standard error in each bin (error bars). Slope uncertainties are bootstrap errors calculated from the unbinned data. A dashed black line at $\Delta$($g-i$) = 0 is included in the lower plot for comparison.}
	\label{res1}
\end{figure}

A significant variation in apparent inclination with total magnitude is apparent ($0.06\pm0.01$), suggesting that brighter galaxies are more likely to be inclined to the line-of-sight. However, this correlation results from the presence of `flagged' galaxies (disc $b/a$ biased by bar component), and the absence of `atypical' galaxies (high $b/a$ on average) towards the more luminous end. Removal of either bias results in no significant variation in axis ratio with luminosity. The apparent inclination of a galaxy does not correlate with its physical properties. Thus, the bias against detection of edge-on systems does not impede the analysis of the colour and luminosity distributions reported in later sections. However, galaxies more inclined than $60^{\circ}$ (b/a $<0.5$) may contain additional undetectable structures (e.g. bars or spiral patterns seen edge-on). As such, these galaxies may yield unreliable model parameters.

\subsection{Bulge and Disc Colours}
\subsubsection{Fixed Multi-band Structure}\label{nograds}
In this section we report on the results where the structural parameters measured in the $g$-band were imposed on the galaxy images in the $u$- and $i$-bands. This results in a uniform colour for both the bulge and disc components at all radii. Thus here, the galaxies' colour gradients are only interpreted as a colour difference between the bulge and disc.

\begin{figure*}
\begin{center}
	\includegraphics[width=\linewidth,clip=true]{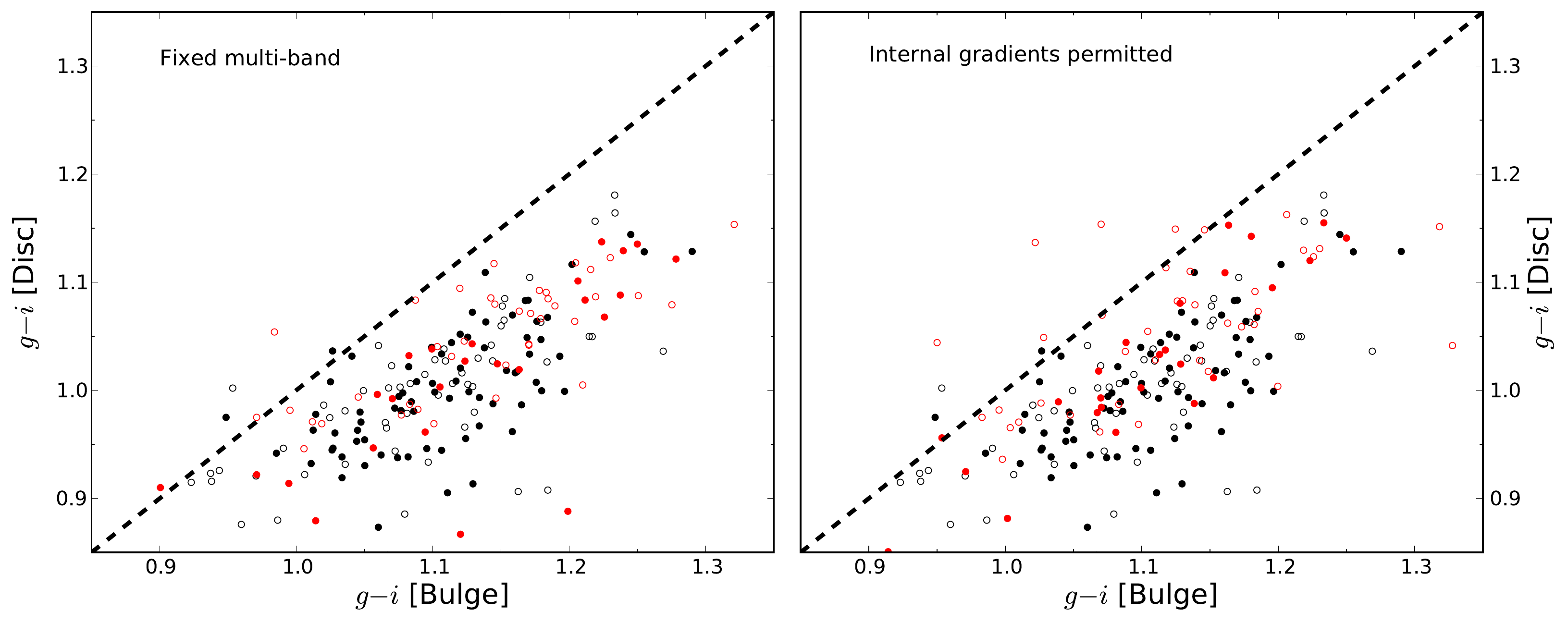}
\end{center}
\caption{Colour-colour ($g-i$) diagrams of analysis sample bulges ($x$-axes) and discs ($y$-axes). {\bf{Left:}} Fixed multi-band models. {\bf{Right:}} Models where internal colour gradients are permitted. Unfilled points indicate `flagged' galaxies, and red points highlight the galaxies which require internal gradients for at least one component. A line of $y=x$ (dashed line) is included in each plot for comparison.}
	\label{CC_nograd}
\end{figure*}

The $g-i$ colours and colour differences between the two components are presented in Figure \ref{res1}. Both the bulges and discs follow significant colour-magnitude trends, with more luminous galaxies possessing both redder bulges and redder discs. This trend is significantly steeper for bulges. A median (bulge $-$ disc) $g-i$ colour offset of $0.097\pm0.004$ mag separates the two components, which increases significantly ($>3\sigma$) for brighter galaxies. For example, an average $M_g = -18$ galaxy has a $g-i$ bulge $-$ disc separation 0.080 mag, while the separation for an average $M_g = -21$ galaxy is 0.146 mag (i.e. $0.066\pm0.0015$ mag larger). The bulge colours have a slightly lower level of scatter in $g-i$ relative to the colours of discs (0.048 and 0.066 mag respectively). 

The $u-g$ colour-magnitude trends are steeper than $g-i$, with no difference between the two components' slopes. Thus, the component colour offset in $u-g$ does not vary significantly with total luminosity. The median value of this offset is $0.178\pm0.007$ mag in $u-g$. Intrinsic scatter of component colours is significantly larger in $u-g$ than $g-i$. For bulges, the scatter relative to the $u-g$ colour magnitude trend is 0.077 mag, while the equivalent scatter for discs is 0.124 mag (i.e. an increase by a factor of $1.6$ and $1.9$ respectively, relative to $g-i$). 

If the colour-magnitude trends are instead measured as a function of component magnitude, the colour-magnitude slopes of bulges and discs are consistent in both $g-i$ (bulges: $-0.046\pm0.003$, discs: $-0.042\pm0.005$) and $u-g$ (bulges: $-0.066\pm0.008$, discs: $-0.083\pm0.010$). The corresponding bulge $-$ disc colour separations at fixed component magnitudes are $\Delta(g-i) = 0.131\pm0.008$ mag and $\Delta(u-g)=0.224\pm0.010$ mag. Here, the colour offsets are measured as a constant offset between the bulge and disc trends, assuming a common slope for the colour-magnitude trends of each component. Thus, $\Delta(g-i)$ and $\Delta(u-g)$ are significantly increased relative to the median (pairwise) separations at fixed total galaxy luminosity reported above.

A colour-colour plot depicting the bulge and disc $g-i$ colours is presented in the left panel of Figure \ref{CC_nograd}. The majority of galaxies form a broad sequence below the illustrative $y=x$ line. This indicates that galaxies with redder bulges have redder discs, and a galaxy's bulge is intrinsically redder than its disc. This is equivalent to the $B-R$ bulge/disc colour plot in \citeauthor{MacArthur2004} (\citeyear{MacArthur2004}; see Figure 8). Converting component colours to $B-R$ ($\approx g-i +0.5$ using \citealp{Jester2005}), the S0s in the present work occupy the same region as the S0/Sa galaxies measured by \citeauthor{MacArthur2004}  

\begin{figure}
\begin{center}
	\includegraphics[width=\linewidth,clip=true]{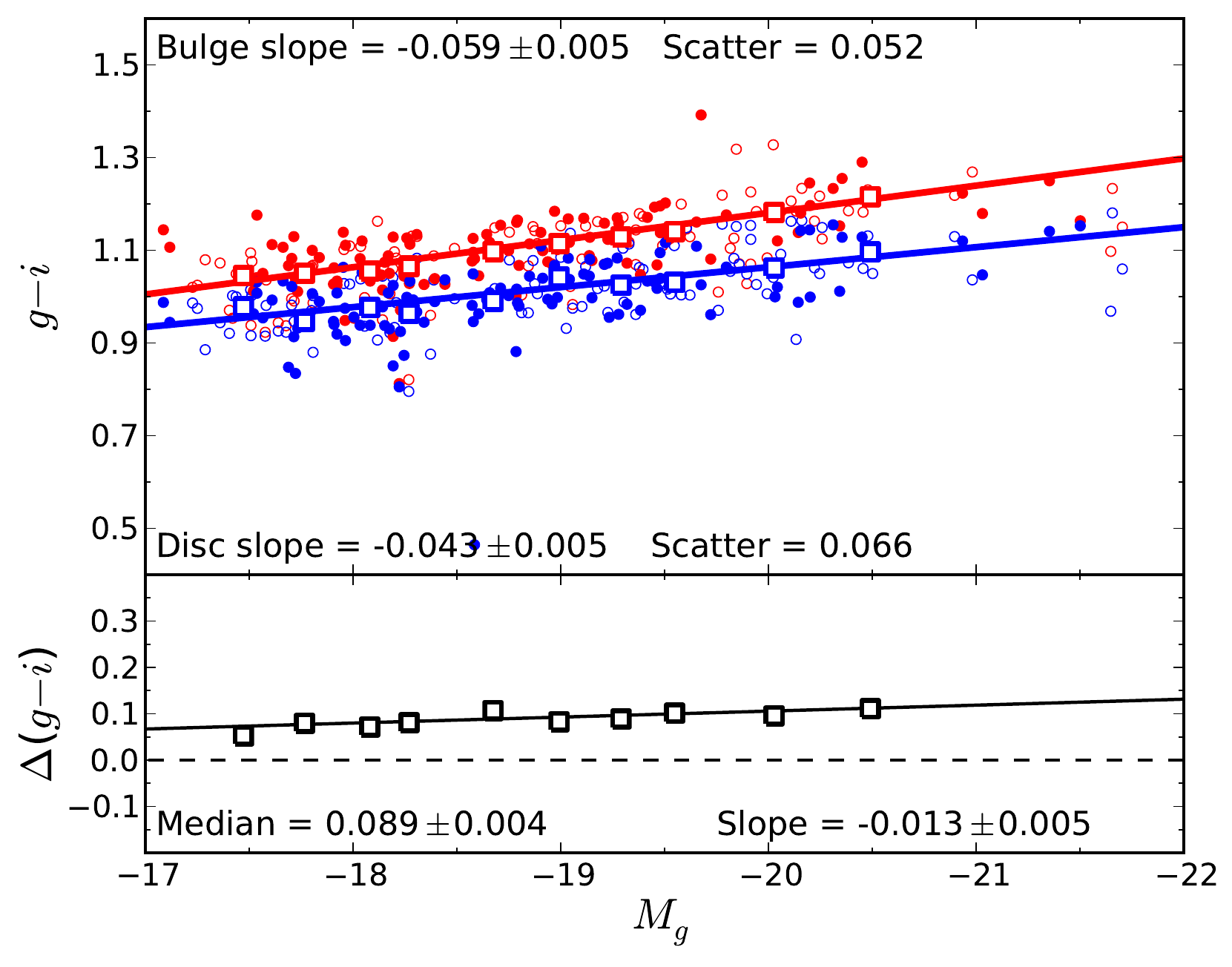}
\end{center}
\caption{As Figure \ref{res1} for $g-i$ colours and $g-i$ colour differences (bulges $-$ discs) when internal component gradients are permitted. 60 galaxies (out of 200) are affected by the inclusion of component gradients.}
	\label{grad1}
\end{figure}

\begin{figure*}
\begin{center}
	\includegraphics[width=\linewidth,clip=true]{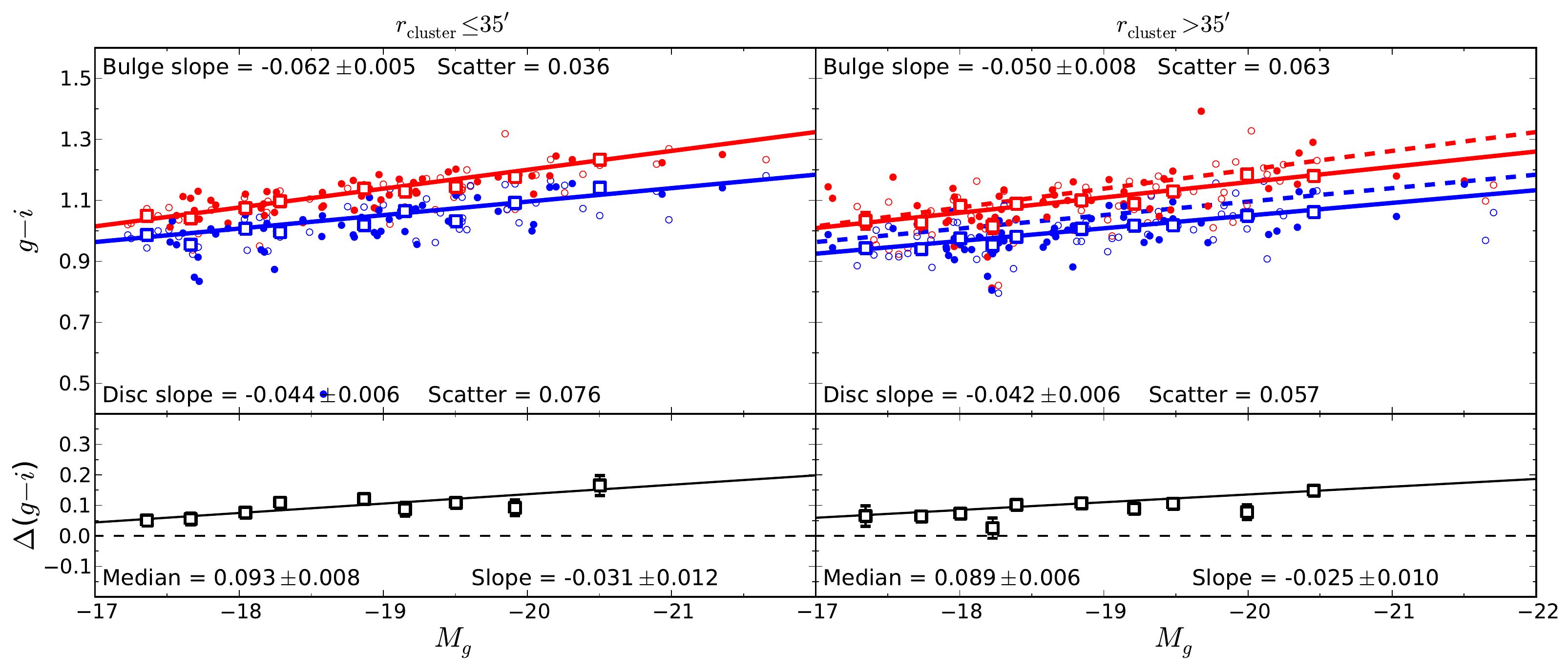}
\end{center}
\caption{Colour-magnitude and colour-separation diagrams (gradients included). The sample is divided at $r_{\rm cluster} = 35\minute$ ($\equiv0.46r_{200}$), defining the inner (left plots) and outer (right plots) cluster samples. {\bf Upper panels:} Bulge (red) and disc (blue) $g-i$ colours plotted against the absolute (bulge + disc) $g$-band model magnitude. {\bf Lower panels:} $g-i$ colour difference (bulge $-$ disc) plotted against  absolute magnitude. Unfilled points indicate `flagged' fit (containing weak extra components or asymmetries, see Appendix \ref{filter}). Best-fit linear trends (solid lines) are fit to median colour values calculated in magnitude bins (large square points), weighted by the robust standard error in each bin (error bars). Slope uncertainties are bootstrap errors calculated from the unbinned data.}
	\label{2rad_CM}
\end{figure*}

\subsubsection{Models with Internal Gradients}\label{grads}
In this section, we describe the results when colour gradients were permitted in the model components to better reproduce the observed colour profiles of the galaxies. Thus here, the sample galaxies' colour gradients are interpreted as a combination of internal colour gradients and a colour difference between the bulge and disc. A full catalogue of resulting (structural and photometric) fitting measurements in the $u$- and $i$-bands is presented in Table \ref{result_2} of Appendix \ref{Cat}.

Of the 200 galaxies in the analysis sample, 60 require component colour gradients (see Section \ref{res_over}). For these galaxies, the average (gradient $-$ no gradient) colour differences for bulge components are $-0.022\pm0.005$ mag and $-0.059\pm0.009$ mag in $g-i$ and $u-g$ (respectively), and $0.009\pm0.005$ mag and $0.023\pm0.011$ mag for discs. Thus, bulges are measured systematically bluer when internal colour gradients are accounted for, while discs are redder (albeit at a marginal level). These systematic colour offsets result in a reduction of the measured bulge $-$ disc colour separation. For the 60 galaxies which require colour gradients, this decreases the apparent median $g-i$ colour separation from $0.104\pm0.010$ mag when internal gradients are ignored, to $0.073\pm0.014$ mag if those gradients are included. In $u-g$, this decrease in colour separation is even more pronounced ($0.175\pm0.013$ without colour gradients; $0.104\pm0.018$ mag with).

The colour shifts measured when gradients were introduced highlights systematic biases in the measurement of bulge and disc colours if internal gradients are not accounted for. Hence, the exaggerated average bulge $-$ disc colour separations for fixed multi-band fitting is a consequence of attempting to account for complex colour profiles with offsets in component colour alone. In all subsequent sections, we report the `best-fit' multi-band model results for the analysis sample ($N=200$). Of these, 60 galaxies include internal gradients in at least one component, while 140 galaxies are adequately fit by a `no component gradient' model.

The $g-i$ colour-magnitude diagram for the analysis sample when internal component gradients are permitted is presented in Figure \ref{grad1}. Here, the median $g-i$ colour separation for the sample as a whole is only $\sim0.01$ mag smaller than was seen previously where component structures were held fixed from band to band. However, colour offsets are most evident towards the bright end of the galaxy magnitude distribution, where component gradients are required more frequently. This results in a larger decrease in the colour separation for brighter galaxies (0.03 mag at $M_g = -21.5$).

As reported above for fixed multi-band fitting, the colour-magnitude slopes as a function of component magnitude for bulges and discs are consistent in $g-i$ (bulges: $-0.045\pm0.003$, discs: $-0.045\pm0.004$) and $u-g$ (bulges: $-0.064\pm0.006$, discs: $-0.075\pm0.007$). Here, the bulge $-$ disc colour offset at fixed component luminosity is $0.123\pm0.008$ mag in $g-i$, and $0.201\pm0.009$ mag in $u-g$.

Allowing for internal component gradients increases the number of free fitting parameters, and hence introduces additional fitting degeneracies. Thus, scatter in measured bulge and disc colours due increases when component gradients are permitted. This effect is illustrated in Figure \ref{CC_nograd} by the increased scatter of gradient-requiring galaxies (red points) in the right panel relative to the left. Nevertheless, this increase in scatter is substantially lower than the corresponding change in average component colours across the analysis sample.

\begin{figure}
\begin{center}
	\includegraphics[width=\linewidth,clip=true]{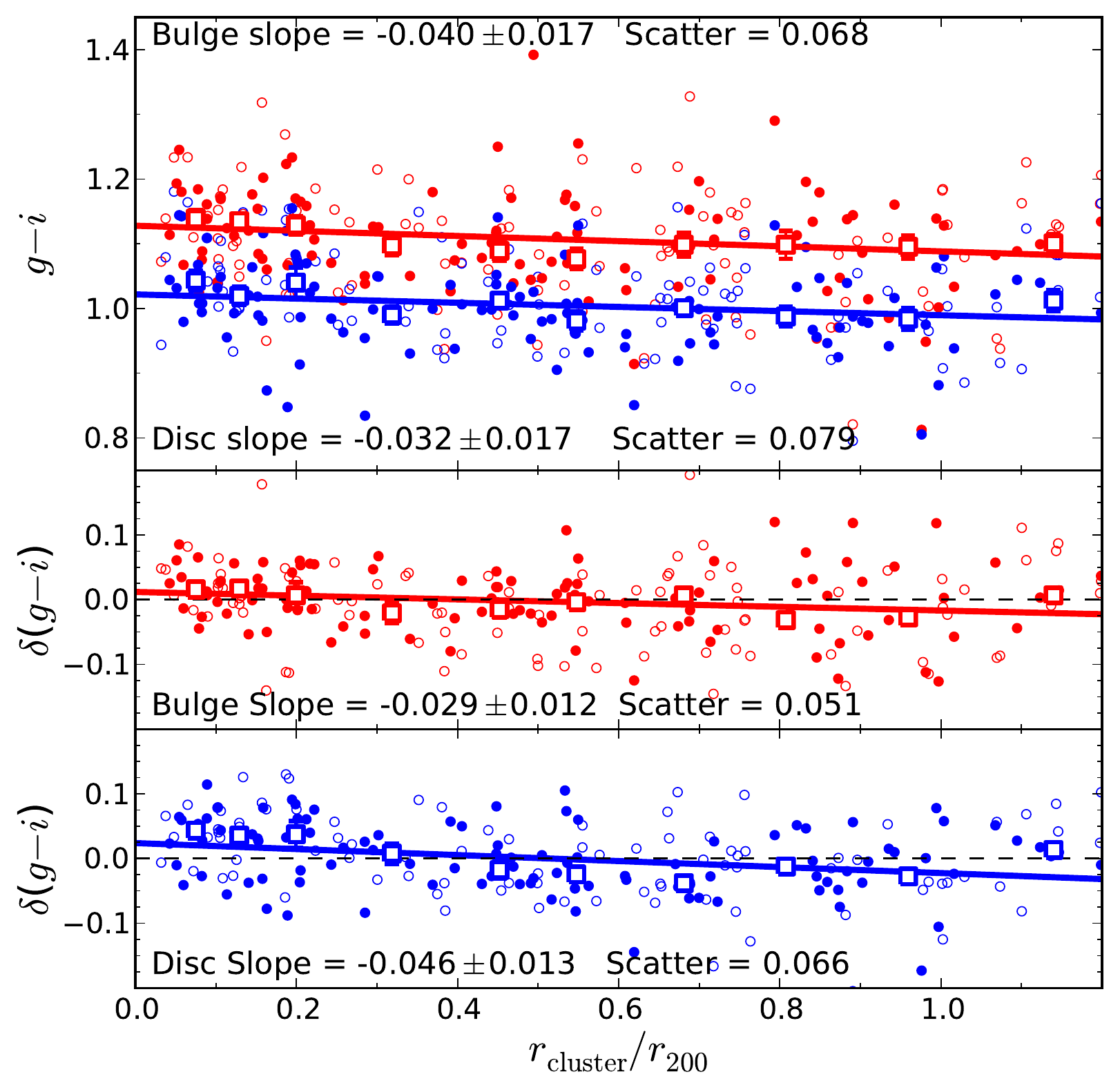}
\end{center}
\caption{{\bf Upper panel:} Colour-radius diagram (component gradients included) displaying the bulge (red) and disc (blue) $g-i$ colours plotted against the projected separation between the galaxy and the Coma cluster centre (in fractional units of $r_{200}$). {\bf Middle Panel:} Bulge $g-i$ colours after subtracting and the equivalent colour-magnitude trend, $\delta(g-i)$, plotted against $r_{\rm{cluster}}$. {\bf Lower Panel:} Disc $g-i$ colours after subtracting and the equivalent colour-magnitude trend, $\delta(g-i)$, plotted against $r_{\rm{cluster}}$. Unfilled points indicate `flagged' fit (containing weak extra components or asymmetries, see Appendix \ref{filter}). Best-fit linear trends (solid lines) are fit to median colour values calculated in bins of projected radius (large square points), weighted by the robust standard error in each bin (error bars). Slope uncertainties are bootstrap errors calculated from the unbinned data. The best-fit bulge and disc trends for the inner sample are included in the upper right plot as dashed lines for comparison.}
	\label{gR1}
\end{figure}

\subsection{Cluster Radial Trends}\label{nograd_rad}
In this section, we examine how the bulge and disc parameters vary as a function of the (projected) distance from the Coma cluster centre, $r_{\rm{cluster}}$. 

With increasing projected distance from the cluster core, galaxies exhibit no significant correlation in bulge \sersic index ($0.12\pm0.29$ $r_{200}^{-1}$), disc axis ratio ($-0.02\pm0.05$ $r_{200}^{-1}$), or the half-light radii of bulge ($-0.12\pm0.45$ $r_{200}^{-1}$) or disc ($0.10\pm0.28$ $r_{200}^{-1}$) components. However, galaxy bulge fraction increases towards the cluster core at a $2\sigma$ level  ($-0.07\pm0.03$ $r_{200}^{-1}$). Thus, while the structure of archetypal bulge + disc galaxies do not change as a function of cluster radius, the relative luminosity of those components varies significantly between the cluster outskirts and core. 

In Figure \ref{2rad_CM}, we sub-divide galaxies into `inner' and `outer' cluster samples based on a cut at $r_{\rm{cluster}}=35^{\prime}$ ($\sim1.0$ Mpc). This radial division is chosen such that each radial sub-sample contains $\sim50\%$ of the analysis sample. Intrinsic scatter in the measured bulge colours of galaxies in the cluster core is remarkably low (0.036 mag) relative to galaxies in the outer sample (0.063 mag). This difference in scatter is significant (from an F-test) at a very high confidence level ($\gg99\%$). The scatter of disc colours in the cluster outskirts (0.057 mag) is comparable to this global colour scatter for all analysis sample galaxies (0.055 mag), but is significantly higher ($>99\%$ confidence) for galaxies in the cluster core (0.076 mag). This increased scatter for discs may result from poorer disc detection at low $r_{\rm cluster}$ due to contamination by neighbouring galaxy haloes. The average bulge $-$ disc colour separations are similar for both samples, but bulge red sequence slopes are weakly ($\lesssim2\sigma$) steeper for galaxies in the cluster core. In addition, we detect a trend towards redder bulges and disc colours at smaller $r_{\rm{cluster}}$. Between the inner and outer samples, the average offset for bulges is $\sim0.03$ mag in $g-i$ and $\sim0.05$ mag in $u-g$. For discs, the offset is $\sim0.04$ mag in both colours.

For the cluster sample as a whole, there exist marginal trends of redder galaxy bulges and discs towards the cluster centre (Figure \ref{gR1}). Discs have a colour-radius slope ($-0.032\pm0.017$ mag $r_{200}^{-1}$ in $g-i$) consistent with the trend for bulges ($-0.040\pm0.017$ mag $r_{200}^{-1}$), but have greater intrinsic scatter (0.079 mag vs. 0.068 mag). However, the core of galaxy clusters are populated by more luminous, bulge-dominated galaxies than the cluster outskirts. Thus, any observed colour-radius trends would need to be decoupled from the colour-magnitude relation. Instead, deviation from the observed (component) colour-magnitude trends, $\delta(g-i)$ and $\delta(u-g)$, as a function of radius is used as an unbiased probe of cluster-radial trends in stellar population properties.

The scatter in $\delta(g-i)$ is $\sim0.02$ mag smaller (relative to the raw colour-radius results) for both components. However, the radial correlation slope is steeper for discs, and shallower for bulges. This change remains small for both components ($\sim1\sigma$), but results in $\delta(g-i)$ slope for discs ($-0.046\pm0.013$ mag $r_{200}^{-1}$) which is significant at a $3.8\sigma$ level. The corresponding $\delta(g-i)$ slope for bulges is weakly significant ($-0.029\pm0.012$ mag $r_{200}^{-1}$), but differs by less than $2\sigma$ from the disc result.

Colour-radius slopes in $u-g$ are significant at a $\sim3\sigma$ level for both components. Compared to $g-i$, these correlations are steeper, and have greater scatter ($\sim1.5\times$). Removing the best-fit component colour-magnitude trends results in shallower $\delta(u-g)$ slopes for both components ($-0.057\pm0.021$ mag $r_{200}^{-1}$ for bulges, $-0.061\pm0.020$ mag $r_{200}^{-1}$ for discs). As with $\delta(g-i)$, these residual trends are consistent but only significant at a $3\sigma$ level for discs. Scatter in $\delta(u-g)$ is $\sim0.02$ mag smaller for both components.

\begin{table*}
\begin{tabular}{lrcccc}
\hline
\hline
\multirow{2}{*}{$N=200$}& & \multicolumn{2}{c}{Fixed Multi-band} & \multicolumn{2}{c}{Internal Gradients} \\
& & Bulge & Disc & Bulge & Disc \\
\hline
\hline
 & $u-g $ & $\mathbf{1.718\pm0.010}$ & $\mathbf{1.520\pm0.010}$ & $\mathbf{1.682\pm0.008}$ & $\mathbf{1.524\pm0.009}$ \\
&  $g-i $ & $\mathbf{1.112\pm0.006}$ & $\mathbf{1.006\pm0.006}$ & $\mathbf{1.108\pm0.006}$ & $\mathbf{1.006\pm0.006}$ \\
Median & $\Delta(u-g)$ [Total]& \multicolumn{2}{c}{$\mathbf{0.178\pm0.007}$} & \multicolumn{2}{c}{$\mathbf{0.164\pm0.007}$} \\
Value & $\Delta(g-i)$ [Total]& \multicolumn{2}{c}{$\mathbf{0.097\pm0.004}$} & \multicolumn{2}{c}{$\mathbf{0.089\pm0.004}$} \\
& $\Delta(u-g)$ [Comp.] & \multicolumn{2}{c}{$\mathbf{0.224\pm0.010}$} & \multicolumn{2}{c}{$\mathbf{0.201\pm0.009}$} \\
& $\Delta(g-i)$ [Comp.] & \multicolumn{2}{c}{$\mathbf{0.131\pm0.008}$} & \multicolumn{2}{c}{$\mathbf{0.123\pm0.008}$} \\
\hline
 & $u-g$ & $\mathbf{-0.078\pm0.008}$ & $\mathbf{-0.078\pm0.010}$ & $\mathbf{-0.085\pm0.007}$ & $\mathbf{-0.077\pm0.007}$ \\
Col-Mag & $g-i$ & $\mathbf{-0.060\pm0.004}$ & $\mathbf{-0.036\pm0.005}$ & $\mathbf{-0.059\pm0.005}$ & $\mathbf{-0.043\pm0.005}$ \\
Slope & $\Delta(u-g)$ [Total] & \multicolumn{2}{c}{$-0.006\pm0.006$} & \multicolumn{2}{c}{$-0.001\pm0.007$} \\
& $\Delta(g-i)$ [Total] & \multicolumn{2}{c}{$\mathbf{-0.022\pm0.005}$} & \multicolumn{2}{c}{$-0.013\pm0.005$} \\
\hline
 & $u-g$ & $-0.039\pm0.032$ & $-0.055\pm0.030$ & $\mathbf{-0.117\pm0.023}$ & $-0.075\pm0.026$ \\
Col-Rad & $\delta(u-g)$ & $-0.010\pm0.022$ & $\mathbf{-0.086\pm0.025}$ & $-0.057\pm0.021$ & $\mathbf{-0.061\pm0.020}$ \\
Slope & $g-i$ & $-0.003\pm0.020$ & $-0.044\pm0.016$ & $-0.040\pm0.017$ & $-0.032\pm0.017$ \\
 & $\delta(g-i)$ & $-0.015\pm0.014$ &$-0.028\pm0.013$ & $-0.029\pm0.012$ & $\mathbf{-0.046\pm0.013}$ \\ 
\hline
\end{tabular}
\caption{Table of median colour values for bulges and discs, and their corresponding colour-magnitude and colour-cluster radius [fractional units of $r_{200}$] trend slopes. $\Delta(C)$ denotes the bulge $-$ disc colour separation measured pairwise as a function of total galaxy luminosity (`Total'), or as a constant offset between bulge and disc trends (sharing a common slope) as a function of component magnitude (`Comp.'). The slope of $\Delta(C)$ [Comp.] with magnitude or radius is 0 by definition. $\delta(C)$ refers to residual colour-radius trends after subtracting component colour-magnitude trends. Results are presented for the entire analysis sample ($N = 200$) for models with no internal gradients (left column) and where internal component gradients are permitted (right column). Significant values ($\geq3\sigma$) are shown in bold.}
\label{nograd_tab}
\end{table*}

\section{Discussion}
In the following sections we summarise the results presented above, and discuss their physical interpretations. First, we examine trends in structural parameters as a means of studying the physical mechanisms which populated the red sequence. Secondly, we connect the measured photometric results to the underlying properties of the galaxies' stellar populations. Lastly, we examine cluster-radial photometric trends in stellar population properties to investigate the effect of local environment on bulge and disc stellar populations. 

As a convenient shorthand, we refer to the red sequence bulge + disc galaxies in the analysis sample as `S0s' hereafter. This is not a rigorous morphological classification, but rather reflects the archetypal central bulge + outer disc galaxy structures selected for analysis in this work.

\subsection{Fingerprints of quenching in structural parameters of S0s}
Different quenching mechanisms (see e.g. \citealp{Boselli2006}) cause distinctive changes to a galaxy's structure in addition to truncating its star formation. As such, the structural ensemble properties of the galaxy sample provide a fossil record of the physical mechanisms which have influenced their evolution. In this section, we investigate trends in bulge and disc structure with the aim of identifying the primary mechanism responsible for the formation of Coma cluster S0s. In particular, we consider the B/T disparity between S0s, and their less bulge-dominated, spiral-like progenitors (\citealp{Dressler1980,Lackner2012}). We discuss this apparent B/T increase in the context of both `bulge-enhancement' and `disc-fading' transformation mechanisms.

In a `bulge-enhancement' scenario, the larger average B/T for S0s results from a build-up of mass (and hence luminosity) in galaxy bulges. Galaxy (major/minor) mergers result in growth of B/T and bulge $n$ \citep{Aguerri2001}, and grow the bulge $R_{e,B}$ proportional to the mass growth, $M^\alpha$ \citep{Naab2009,Hilz2012}. Tidal torques induced by mergers also drives disc gas towards the galaxy centre, which may quench the star formation via rapid consumption of its gas reservoir and/or AGN feedback. If brighter (higher B/T) galaxies built a greater fraction of their present-day mass through merger events, then bulge $R_{e,B}$ should correlate with total galaxy luminosity. 

Major merger (mass ratios 1:1 to 3:1) result in an approximately linear growth in bulge size relative to mass growth ($\alpha = 0.8-1.0$; \citealp{BKolchin2005,Hilz2012}). In addition, disc structures are disrupted or destroyed by major mergers \citep{Barnes1991,Bendo2000,Bois2011}, yielding a (kinematically) elliptical-like remnant. These predictions are incompatible with our observed size-luminosity relations; the measured physical sizes of these bulges do not vary significantly with luminosity over the galaxy luminosity range analysed in the present work (Figure \ref{g_structure}c). A correlation may exist for more luminous galaxies ($M_g < -20.5$), however the small number of galaxies in this luminosity range limits its detection in the present sample. For galaxy discs, we measure a strong correlation between half-light radius and total galaxy luminosity, indicating that the discs survive after transformation. This trend is present even if the requirement that discs dominate the outer part of their galaxies (i.e. archetypal structures) is relaxed, and thus is not a consequence of our selection bias. Furthermore, when compared to the discs of local, star-forming spiral galaxies, the S0s in Coma have a similar size-luminosity slope (Figure \ref{siz_lum}), albeit with an offset in magnitude (see disc fading discussion below). Thus, the structural scaling relations of (spiral-like) progenitor galaxies are preserved for S0 discs. We therefore rule out disc-destructive mechanisms (i.e. major mergers) as the dominant process in S0 formation. 

Minor mergers (mass ratios 5:1 to 10:1) provide a less disc-disruptive means of increasing B/T \citep{Walker1996}. Remnant disc structures are thickened \citep{Quinn1993}, but retain the kinematic and (radial) size properties of their progenitors \citep{Bournaud2005}. Conversely, a sequence of multiple minor merger events could cumulatively destroy disc structures \citep{Bournaud2007}, yielding remnants similar to a major merger. Bulge size growth is predicted to be significantly larger in minor mergers than for lower mass-ratios ($\alpha=2.0-2.4$; \citealp{vDokkum2010,Hilz2012}), even more inconsistent with the flat bulge size-luminosity trend detected in the present work for intermediate/faint S0s. However, not all simulations agree on this point. For example, \cite{EMoral2012} suggest that dry, minor mergers can provide the bulge growth necessary for S0 formation while causing no significant change to either bulge or disc scale lengths. If true, a minor merger scenario would be in agreement with our results, provided that component sizes can be preserved over multiple minor merger events.

In the context of Coma cluster S0s, direct mergers are unlikely due to the large relative velocities involved in the cluster galaxy interactions. Hence, any merger-driven transformation more likely reflects `pre-processing' (e.g \citealp{Wilman2009}) of S0 progenitors at early times, rather than evolution mediated by the cluster environment. Instead, a sequence of rapid tidal interactions  (`galaxy harassment'; \citealp{Moore1996}) may drive S0 formation; quenching star-formation \citep{HFernandez2012}, while thickening \citep{Moore1999} and fading \citep{Moore1998} progenitor discs. A generic `disc fading' mechanism (also e.g. ram pressure stripping or strangulation) can thus explain the B/T increase between S0s and spirals. 

\cite{Rawle2013} found that Coma S0 galaxies are offset from the Tully-Fisher Relation (TFR) for spirals such that for a fixed rotation velocity, S0s are $1.1\pm0.2$ mag fainter in total galaxy luminosity than spirals in the $g$-band. Assuming that this offset is a pure change in disc luminosity, then the progenitor of an `average'  Coma S0 ($M_g =-18.8\pm0.1$ mag, B/T $=0.38\pm0.01$) was brighter ($M_{g,0}=-19.9\pm0.2$ mag) and less bulge dominated (B/T$_0 = 0.14\pm0.02$). Hence, disc components were faded by $1.5\pm0.2$ mag on average in $g$. However, we have shown that for a fixed disc size, Coma S0s are {\emph{brighter}} ($g$-band: $0.52\pm0.20$ mag globally, $0.27\pm0.12$ mag for discs) than local star-forming spiral galaxies (Figure \ref{siz_lum}). Thus, the simplest model of disc fading - wherein only disc {\emph{luminosity}} is changed - is incompatible with the results of the present work.

To reconcile disc fading with the observed size-luminosity relations, either progenitor discs must be intrinsically smaller (and/or brighter) than the discs of present-day spirals, or discs must have been truncated in size during transformation to S0s (as previously noted in Coma; \citealp{Aguerri2004}). This may be observed as e.g. `type II' surface brightness profiles (see \citealp{Pohlen2006,Roediger2012}). Physical truncation may result from the stripping of outer disc stars predicted from galaxy harassment \citep{dJong2004,Aguerri2009}. Fading the total galaxy luminosity of star-forming spirals by 1.1 mag requires a disc size offset (i.e. S0 discs smaller than spiral discs) of $1.5\pm0.3$ kpc (at fixed galaxy luminosity) to be consistent with the observed global Coma S0 size-luminosity trend. Repeating this analysis for size-luminosity as a function of disc luminosity yields a similar size offset ($1.5\pm0.2$ kpc). From the average observed disc size for our sample ($R_s =2.2\pm0.1$ kpc), S0 discs are thus $41\pm6\%$ smaller on average than the equivalent discs of local spiral galaxies. 

A disc-fading scenario is supported by the detection of a strong correlation between disc component and total galaxy colours \citep{Lackner2012}\footnote{\citeauthor{Lackner2012} also report an increase in ``bulge size" (from B/T) as galaxies transition onto the red sequence. However, it is not necessary to invoke bulge growth to explain an increase in B/T if disc fading is a dominant effect.}. The (global) colour transition from blue [cloud] to red [sequence] galaxies was thus interpreted as a driven by redder disc colours. Disc-fading was disputed by \citeauthor{Burstein2005} (\citeyear{Burstein2005}; see also \citealp{Christlein2004}), who found that S0s have equal (or greater) $K$-band luminosity than spirals, rather than less. However, this offset compares present-day S0s to present-day spirals, rather than to their progenitors. In addition, the measurement of {\emph{global}} luminosity differences does not strictly rule out disc fading (i.e. if the bulge is brighter).

In summary, through analysis of the structural parameters of the bulges and discs of S0s, we have investigated the physical mechanisms responsible for their formation. The detection of a strong size-luminosity relation for discs - but not for bulges - does not support a bulge-enhancing origin for S0s, particularly for disc-disruptive (major) and bulge size growing (minor) mergers. While the probable importance of mergers in bulge mass assembly at early times has been previously noted (e.g. \citealp{Fisher1996,Barway2009,Barway2013}), here we are asserting that they are not responsible for the final transformation which yields the observed S0s. This work therefore supports disc-fading as the dominant formation mechanism for S0s in the Coma cluster. However, as S0s are brighter (at a fixed disc size) than local spirals, S0 discs must be significantly smaller than the discs of today's star-forming spirals. Therefore, either the discs of cluster S0 progenitors were intrinsically smaller or brighter than local field spiral discs, or these discs were truncated during transformation to S0s. A progenitor size bias is in qualitatively agreement with \cite{Trujillo2006}, who found low $n$ objects at $z\sim2.5$ $\sim3\times$ smaller than local galaxies. From their size growth predictions, we estimate a corresponding quenching redshift of $z\sim1.5$.

\subsection{The component red sequences of S0 bulges and discs}
In this section we consider the separated photometry of bulges and discs in our archetypal S0 sample. Through analysis of component colour-magnitude trends, we investigate whether Coma's S0 red sequence is a consequence of increasingly red colours in only one (e.g. bulges in H10), or both components for more luminous galaxies. As such, we place photometric constraints on whether quenching acts primarily on one component.

The slope of the red sequence implies a luminosity dependence of the average stellar population properties (age and/or metallicity). In H10, this global colour-magnitude slope was found to be a consequence of change in bulge colour in cluster galaxies. For the Coma S0s in this work, we measure a bulge colour-magnitude slope of $-0.059\pm0.005$ (Figure \ref{grad1}). This is consistent with the equivalent $B-R$ slope measured in H10 ($-0.037\pm0.014$). However, we {\emph{also}} measure a similar significant colour-magnitude slope for S0 discs ($-0.043\pm0.005$). By contrast, H10 reported no colour-magnitude slope for disc components ($0.000\pm0.010$). This difference is not a consequence of selection bias: the colour-magnitude slopes measured in this work are not significantly altered if either the initial colour cut ($g-r > 0.5$) or later sample vetting (see Section \ref{decomp} and Appendix \ref{filter}) are removed. Thus, we find that the red sequence of our archetypal S0 sample as due to similar trends towards redder colours for both structural components, rather than being driven only by the bulge colour-magnitude relation. 

The colours of bulges and discs represent a fossil record of their stellar populations' star-formation history and subsequent quenching. A $\Delta(B-R)$ colour separation ($\simeq \Delta(g-i)$) of $\sim0.25$ between bulges and discs of $L_*$ galaxies was previously reported in H10. Since most galaxies have a negative colour gradient, this colour offset reflects the radial positions of the bulge and disc structures in S0s. As such, bulge stellar populations are expected to be older (or more metal-rich) than discs. 

The bulges of Coma S0s are measured in this work to be $0.089\pm0.004$ mag redder than their discs on average in $g-i$, and $0.164\pm0.007$ mag redder on average in $u-g$ at fixed total galaxy luminosity. At a fixed component luminosity, bulges are $0.123\pm0.008$ mag redder than discs on average in $g-i$, and $0.201\pm0.009$ mag redder on average in $u-g$. Hence, bulges are significantly redder than discs in both colours when either measured between components of an average galaxy, or between equal luminosity components. 

The average $\Delta(g-i)$ colour separation reported here at fixed total luminosity is thus $\sim3\times$ smaller than the equivalent $\Delta(B-R)$ offset reported in H10, while the offset at fixed component luminosity is $\sim2\times$ smaller. Removing the sample filters yields a significantly bluer median colour for galaxy bulges, although the median disc colours remain consistent with the filtered sample median. However, the average colour offsets are not changed significantly. Thus, the measurement of a smaller bulge $-$ disc colour offsets for red sequence galaxies relative to H10 is a secure result, robust to the choice of sample selection criteria. 

Simple stellar population (SSP) models connect measured photometry to the underlying stellar population properties of a galaxy. Applying multi-linear fitting to \cite{Maraston2005} SSP models (single burst star formation, Kroupa initial mass function) yields expressions for $g-i$ and $u-g$ colours as linear combinations of stellar population ages and metallicities. The bulge and disc colours, and colour separations measured above can thus be used to constrain the timescales involved in galaxy quenching. However, given the degenerate effects of age and metallicity on optical colours \citep{AgeMet}, we do not attempt to attribute colour differences to exact values of either stellar population property.

If interpreted as solely an age difference, the $0.089\pm0.004$ mag bulge $-$ disc colour separation in $g-i$ (at a fixed galaxy magnitude) corresponds to a bulge age ($t_B$)  $(1.9\pm0.1)\times$ larger than the disc age ($t_D$) on average. Using the $u-g$ colour separation, a larger age offset of $t_B = (3.0\pm0.5)t_D$ is measured. To estimate the absolute age difference between bulges and discs, $\Delta t$, we assume $t_B = t_{\rm max} = 13.8$ Gyr. Using $\Delta (g-i)$, S0 bulges are thus $6.4\pm0.4$ Gyr older than their discs on average. This is equivalent to discs formation at $z_{\rm{disc}} = 0.9\pm0.1$. From $\Delta(u-g)$, a significantly larger age offset, $\Delta t = 9.2\pm0.8$ Gyr ($z_{\rm{disc}}=0.4\pm0.1$) is measured between the two components. The apparent discrepancy between these estimates is caused by increased sensitivity of $u-g$ to emission from young stars (i.e. discs are younger than detected in $g-i$). These disc ages are consistent with the presence of recent star formation activity in a fraction of the measured Coma S0s. This may indicate contamination of the sample by red spiral galaxies which have been dust-reddened (e.g. \citealp{Cibinel2013}), or are currently undergoing truncation (e.g. \citealp{Crossett2013}). 

Alternatively, if interpreted as only a difference in the metallicity, the ($g-i$) colour separation between bulges and discs corresponds to a bulge metallicity ($Z_B$) $(2.2\pm0.1)\times$ greater than the metallicity of the disc ($Z_D$). From $u-g$, the metallicity difference would be $Z_B = (2.4\pm0.1)Z_D$ for fixed galaxy magnitude. 

For typical ETGs, spectroscopic investigation of stellar populations report negative global metallicity gradients \citep{Fisher1996,Moorthy2006,Morelli2008,Rawle2010,Johnston2012,LaBarbera2012} indicating galaxy centres more metal-rich than than their outer regions. Global age gradients have been reported as either weakly negative/flat \citep{Kuntschner2010,Rawle2010,Eigenthaler2013}, or positive \citep{Bedregal2011,Johnston2012,LaBarbera2012}. If both age and metallicity gradients are negative, then the negative colour gradients measured in this work correspond to bulge $-$ disc age/metallicity offsets smaller than calculated above. Conversely, if age gradients are positive (i.e. centres of S0s younger than their outer regions), then a larger metallicity offset is required, relative to the (pure metallicity) values calculated above.

The measured bulge $-$ disc colour offsets can be interpreted as the difference in stellar population properties between the half light radii of the bulge and disc (with $R_{e,B} < R_{e,D}$ in archetypal S0s). On average, this corresponds a $0.089$ mag decrease in $g-i$ colour (0.164 mag decrease in $u-g$) across $\Delta {\rm log}(R) = -0.55$ dex in radius. A negative metallicity gradient and a weakly {\emph{negative}} age gradient ($\Delta[Z/H]/\Delta{\rm log}(R)=-0.13\pm0.04$ and $\Delta {\rm log(age)}/\Delta {\rm log}(R) = -0.02\pm0.06$ from \citealp{Rawle2010}) yield bulge $-$ colour offsets of $\Delta(g-i)=0.022\pm0.013$ mag and $\Delta(u-g)=0.036\pm0.017$ mag. These predicted colour offsets are significantly smaller than the observed $\Delta(g-i)$ and $\Delta(u-g)$, indicating that stellar population gradients alone are insufficient to explain the colour difference between bulges and discs.  Further, the \cite{Rawle2010} sample is dominated by E or E/S0 morphologies. The metallicity gradient for a sample of S0s would be shallower (e.g. \citealp{Koleva2011}), and thus the gradient-predicted colour offset would be even more discrepant with the measured values presented above. Alternatively, a negative metallicity gradient and positive age gradient ($\Delta[Z/H]/\Delta{\rm log}(R)=-0.6\pm0.5$ and $\Delta {\rm age}/\Delta {\rm log}(R) = 2.3\pm4.6$ Gyr dex$^{-1}$ from \citealp{Chamberlain2011}) yield bulge $-$ colour offsets of $\Delta(g-i)=0.097\pm0.074$ mag and $\Delta(u-g)=0.160\pm0.102$ mag, assuming a bulge age of $t_B =13.8$ Gyr. These colour offsets are similar to the observed values, but the substantial uncertainty prevents a strong conclusion from being drawn.  

To summarise, the colour-magnitude slopes of S0 bulges and discs are similar, with an average colour separation of $\sim0.1$ mag in $g-i$, and $\sim0.2$ mag in $u-g$. Using SSP models, the redder colours of S0 bulges can be calculated as a difference in either a stellar population age, or metallicity relative to S0 discs. At a fixed galaxy luminosity, bulges are found to be $\sim2$-$3\times$ older than discs, or $\sim2\times$ more metal rich. 

\subsection{The effect of environment on bulge and disc colours}\label{Summ2}
Here, we discuss variations of component colours with projected distance from the cluster centre. Observed $r_{\rm cluster}$ correlates with the time at which a galaxy first entered the cluster environment \citep{Gao2004,Smith2012,DeLucia2012,Taranu2012}, albeit with substantial scatter. A radial analysis therefore highlights variations of stellar population properties during cluster infall. By investigating the cluster-centric radial trends for bulges and discs, we thus trace the environment-mediated processes that have acted on these structural components, and hence the cluster environment's role in star formation quenching and S0 formation. 

The best-fit component colour-magnitude trends are subtracted to control for the trend towards brighter, more bulge-dominated galaxies at lower $r_{\rm{cluster}}$. The resulting residual colour trends ($\delta(C)$, where $C$ is either $g-i$ or $u-g$) with $r_{\rm cluster}$ therefore provide estimates for radial trends in stellar population properties independent of this luminosity bias. Radially outwards from the cluster core, S0 discs become significantly bluer ($-0.046\pm0.013$ mag $r_{200}^{-1}$) in $g-i$ (Figure \ref{gR1}). S0 bulge $g-i$ colour, however, only correlates weakly with $r_{\rm cluster}$ ($-0.029\pm0.012$ mag $r_{200}^{-1}$). Steeper colour-radial trends are measured in $u-g$ for both components, but at similar significance levels to $g-i$. These results are unchanged if instead multilinear fitting is used to investigate component colour as a function of component magnitude and cluster radius simultaneously.

H10 previously noted a significant dependence of disc colours on projected cluster-centric radius ($-0.103\pm0.023$ mag $r_{200}^{-1}$ in $B-R$). This trend is steeper, but consistent at a $\sim2\sigma$ level with the correlation reported above for $g-i$ disc colours. Similarly, H10 bulge colours were found to be invariant with radius ($0.007\pm0.038$ mag $r_{200}^{-1}$ in $B-R$), consistent with the (weak) correlation between $g-i$ bulge colour and $r_{\rm cluster}$ in this work. Conversely, \cite{Lackner2013} found that disc colours drove global colour-density trends, but only for galaxies in poor groups. In rich clusters, disc colours were measured as invariant with local density, while bulge colours were weakly anti-correlated with environment.

The measured $\delta(C)$-radius slopes are equivalent to colour offsets between galaxies of similar luminosity at $r_{\rm cluster} = 0$ and $r_{\rm cluster} = r_{200}$. Thus, the residual bulge and disc slopes can be used to compare their stellar population ages and metallicities at Coma's virial radius ($t_{200}, Z_{200}$) to those in the core ($t_0,Z_0$). Interpreted as a pure age difference in S0 discs with radius (via SSP models as above), the $\delta(u-g)$ slope of $-0.061\pm0.020$ mag per $r_{200}$ yields $t_{\rm 0,D}=(1.5\pm0.2)t_{\rm 200,D}$. Equivalently from the $\delta(g-i)$ slope, the disc age difference is estimated as $t_{\rm 0,D}=(1.4\pm0.1)t_{\rm 200,D}$. The stellar population age differences for S0 bulges are $t_{\rm 0,B}=(1.5\pm0.2)t_{\rm 200,B}$ from $\delta(u-g)$, and $t_{\rm 0,B}=(1.2\pm0.1)t_{\rm 200,B}$ from $\delta(g-i)$. Upper limits can be placed on bulge age offsets by using $t_{0,B} = t_{\rm max} = 13.8$ Gyr. From $\delta(u-g)$, S0 bulges in Coma's core are older than those found in the cluster outskirts by $4.3\pm1.4$ Gyr, while the equivalent age difference from $\delta(g-i)$ is $2.5\pm0.9$ Gyr. No corresponding limits can be placed on the ages of S0 discs without an estimate for $t_{\rm max,D}$. 

Alternatively, if a pure metallicity difference was responsible for the $\delta(u-g)$ component colour changes between the virial radius and cluster core, then $Z_{\rm C,B}=(1.36\pm0.15)Z_{\rm 200,B}$ for bulges, and $Z_{\rm C,D}=(1.39\pm0.15)Z_{\rm 200,D}$ for discs. From $\delta(g-i)$, the equivalent metallicity changes would be $Z_{\rm C,B}=(1.30\pm0.14)Z_{\rm 200,B}$ and $Z_{\rm C,D}=(1.51\pm0.18)Z_{\rm 200,D}$.

In summary, we confirm that the increasingly red total colours of S0 galaxies towards the cluster centre are mainly driven by a cluster-radial trend in the colours of component discs. The detection of a weak trend in bulge colours with radius indicates that while environment-driven bulge contributes to S0 formation, it is not a dominant factor. Therefore, our component colour data favour S0 formation via an environment-driven fading of the discs of spiral-like progenitors. Thus, information about the earlier formation history of S0 bulges is largely preserved in their stellar population properties.

A comparison of simple quenching model scenarios (e.g. \citealp{Smith2012, Taranu2012}) with the observed colour differences in this work is necessary to fully understand the quenching histories of red sequence galaxies (e.g. can the observed colour scatter be explained by instantaneous quenching upon cluster infall?). This will be explored in depth in a later paper.

\section{Conclusions}
We have explored the structure and photometry of red-sequence galaxies in the Coma cluster using deep $u$-, $g$-, and $i$-band imaging. The initial sample comprised 574 Coma cluster member galaxies (in the cluster-radial range $0<r_{\rm{cluster}}<1.3$ $r_{200}$) with $M_g < -17$ and $(g-r)>0.5$. Using GALFIT, a \sersic + exponential model was fit to all galaxies in the sample, allowing their constituent bulge and disc structures to be investigated separately. 

Extensive sample vetting was carried out to isolate a sample of well-fit, 2-component galaxies for which the assumption of `archetypal' (central) bulge + (outer) disc morphology is valid. The fitting solution was considered to be `stable' (i.e. symmetric, uncontaminated, with $\chi^2\sim1$) in 443 galaxies ($77\%$) of the initial sample. A significant additional structural component was detected in 337 ($76\%$) of these stable fits. The bulge-disc structure was deemed `atypical' (the bulge does not dominate {\emph{only}} at the galaxy's centre and/or the disc does not {\emph{only}} dominate its outer regions) in 137 galaxies ($41\%$ of 2-component fits). The remaining 200 bulge+disc galaxies ($59\%$) were considered archetypal S0s, and defined our analysis sample (see Figure \ref{Venn_samp}).

Fitting was carried out using two approaches: i) where model structure does not vary from band-to-band (``fixed multi-band'' fitting); ii) where bulge $R_e$ and/or disc $R_s$ is allowed to vary in order to incorporate internal component colour gradients. The latter approach provides a significant improvement to the colour profile fitting for $30\%$ of the analysis sample galaxies. The fraction of galaxies requiring component gradients decreases with luminosity, suggesting that the detection of internal component gradients is magnitude limited. Measured component colours, and trends with galaxy luminosity and cluster-centric radius are summarised in Table \ref{nograd_tab} for both multi-band fitting techniques. 

Our main results are as follows:

\begin{enumerate}[label=\roman{*}),ref=(\roman{*})]
\item S0 bulges have physical sizes, $R_{e,B} \sim 1$ kpc, and profile shape, $n\sim2$, on average. Bulge are thus structurally similar to both dwarf ellipticals in Coma, and ``red nugget" galaxies observed at $z > 1.5$ (albeit an order of magnitude smaller in mass than the latter). 
\item Disc effective half-light radius is strongly correlated with total galaxy luminosity. The slope for S0 discs is consistent with the size-luminosity relation for today's star-forming disc galaxies. However, these trends are separated in luminosity such that S0s are brighter than such spirals. If S0 galaxies are formed through disc fading, then their discs are $40\%$ smaller on average than the discs of equally luminous local spirals. Thus, either their progenitor's discs were intrinsically smaller than the discs of today's spirals (suggesting a quenching epoch of $z\sim1.5$), or were truncated in size during transformation to S0s.
\item No significant correlation between bulge half-light radius and total galaxy luminosity is detected (for $M_g > -20.5$). This is inconsistent with predictions for a `bulge-enhancement' scenario of S0 formation via major mergers ($R_e$ increases $\propto M$) or minor mergers ($R_e \propto M^2$). 
\item If either galaxy or component luminosity is fixed, the bulges of S0s are significantly redder than their discs (by $\sim0.1$ mag in $g-i$, $\sim0.2$ mag in $u-g$ on average). Thus, bulge stellar populations are significantly older, and/or more metal rich than those found in discs.
\item Significant colour-magnitude slopes are detected for both structural components in $g-i$ and $u-g$. In either colour, the measured trend slopes for bulges and discs are consistent at a $2\sigma$ level. Hence, the global red sequence is a consequence of the increasingly red colours of bulges {\emph{and}} discs in more luminous galaxies.
\item After subtracting the best-fit colour-magnitude trend, galaxy discs become bluer in both $g-i$ and $u-g$ with increasing projected distance from the centre of the Coma cluster. Bulge component colours also become bluer with $r_{\rm cluster}$, but this trend is only marginally significant ($2.5\sigma$) and more sensitive to the treatment of component colour gradients. The global colour-radius trend for S0s is thus dominated by increasingly red discs in galaxies closer to the cluster core. Therefore, the environment-mediated mechanism which drives S0 formation is a `disc-fading' process.
\end{enumerate}

These results are not significantly changed by either including the blue galaxies removed during the selection of the initial galaxy sample, or the removal of the {\it a posteriori} sample vetting used to select only well-fit, `archetypal' galaxies.

From the results of fitting `archetypal' bulge + disc galaxies, we interpret the red sequence as a consistent shift towards redder colours for both the bulges and discs of more luminous galaxies. Significant trends towards bluer discs (but only marginally bluer bulges) further from the cluster core indicate that the colour-environment relation is caused by an environment-driven disc fading mechanism. To reconcile disc fading with more luminous S0s, either the discs of S0s must be truncated size during transformation, or their progenitors were intrinsically smaller and/or fainter than today's star-forming spirals.

As our study focuses purely on galaxies with `archetypal' bulge + disc structures, we do not rule out a merger-based formation scenario in {\emph{all}} early-types. Our `atypical' galaxies represent the most bulge-dominated ETGs, and hence are the most likely candidates for `traditional' ellipticals. The possibility of multiple evolutionary pathways for `archetypal' and `atypical' ETGs (e.g. due to a relatively quiet merger history for `archetypal' S0s) will be the subject of a future study.

\section*{Acknowledgements}
We thank the anonymous referee for their comments, which have improved the paper. We also thank Mark Norris for helpful comments.
JTCGH is supported by an STFC studentship (ST/I505656/1). JRL and RJS are supported by STFC Rolling Grant ST/I001573/1. MJH acknowledges support from NSERC (Canada). 

This work is based on observations obtained with MegaPrime/MegaCam, a joint project of CFHT and CEA/DAPNIA, at the Canada-France-Hawaii Telescope (CFHT) which is operated by the National Research Council (NRC) of Canada, the Institute National des Sciences de l'Univers of the Centre National de la Recherche Scientifique of France, and the University of Hawaii. This work is based in part on data products produced at TERAPIX with the expert assistance of Partick Hudelot and Yannick Mellier. Observational data used in this paper are available from the CFHT archive http://www3.cadc-ccda.hia-iha.nrc-cnrc.gc.ca/cfht/cfht.html

This work uses data from SDSS-III. Funding for SDSS-III has been provided by the Alfred P. Sloan Foundation, the Participating Institutions, the National Science Foundation, and the U.S. Department of Energy Office of Science. The SDSS-III web site is http://www.sdss3.org/.

SDSS-III is managed by the Astrophysical Research Consortium for the Participating Institutions of the SDSS-III Collaboration including the University of Arizona, the Brazilian Participation Group, Brookhaven National Laboratory, Carnegie Mellon University, University of Florida, the French Participation Group, the German Participation Group, Harvard University, the Instituto de Astrofisica de Canarias, the Michigan State/Notre Dame/JINA Participation Group, Johns Hopkins University, Lawrence Berkeley National Laboratory, Max Planck Institute for Astrophysics, Max Planck Institute for Extraterrestrial Physics, New Mexico State University, New York University, Ohio State University, Pennsylvania State University, University of Portsmouth, Princeton University, the Spanish Participation Group, University of Tokyo, University of Utah, Vanderbilt University, University of Virginia, University of Washington, and Yale University. 

\bibliographystyle{mn2e}
\bibliography{jtcgh.bib}

\appendix
\section{Automated GALFITting of Optical and Near-Infrared Imaging (AGONII)}\label{AGONII}
AGONII is a python wrap-around script for GALFIT developed for this work to extend the fitting process, providing a more thorough search of $\chi^2$-space. This circumvents the two major issues with running stand-alone GALFIT: Sensitivity of fitting results to initial parameters, and inability to differentiate between local and global $\chi^2$ minima.  In addition, AGONII fully automates the bulge-disc decomposition analysis presented in this work, including multi-band fitting, and sample filtering. 

AGONII's fitting extension is implemented by perturbing the GALFIT model parameters from their apparent best-fit positions and re-fitting. A large perturbation (a $3\times$ increase/decrease) will shift GALFIT out of any local minimum in $\chi^2$-space, thus enabling the true global minimum to be detected. This process is repeated a minimum of twice per model parameter (e.g. brightness, size and axis ratio) until a stable, global $\chi^2$ minimum solution is found.

In order to acquire realistic starting values for our fit parameters, we build up model complexity over subsequent stages of the analysis. We initially exhaustively fit a single \sersic model to the galaxy. The resulting best-fit is then used as the basis for the 2-component fit initial conditions. Generic initial conditions are used for this one-component model. The choice of these starting values has no effect on the final, two-component results due to the low level of degeneracy and scatter in a single-component fit. The \sersic component is duplicated (increasing both components' magnitudes accordingly). The duplicate component has its \sersic index fixed at $n = 1$ (i.e. an exponential disk). We run GALFIT again on this input \sersic + exponential model, and the resulting fit parameters are used as the initial values for the two-component fitting.

The influence of non-physical models can be mitigated through use of fitting constraints to set upper and lower limits to the acceptable parameter values. Excessive use of limits can introduce model biases, however, preventing GALFIT from finding the best-fit model. In this study, constraints are only used to provide extreme limits to model parameters. This minimally-constrained approach reduces the likelihood of overflow errors or non-physical parameter values. Instead, a thorough {\it a posteriori} sample filter (as described in Appendix \ref{filter}) is utilised to ensure reliable fitting results. 

\section{Sample Filtering}\label{filter}
\begin{figure}
\begin{center}
\includegraphics[width=\linewidth]{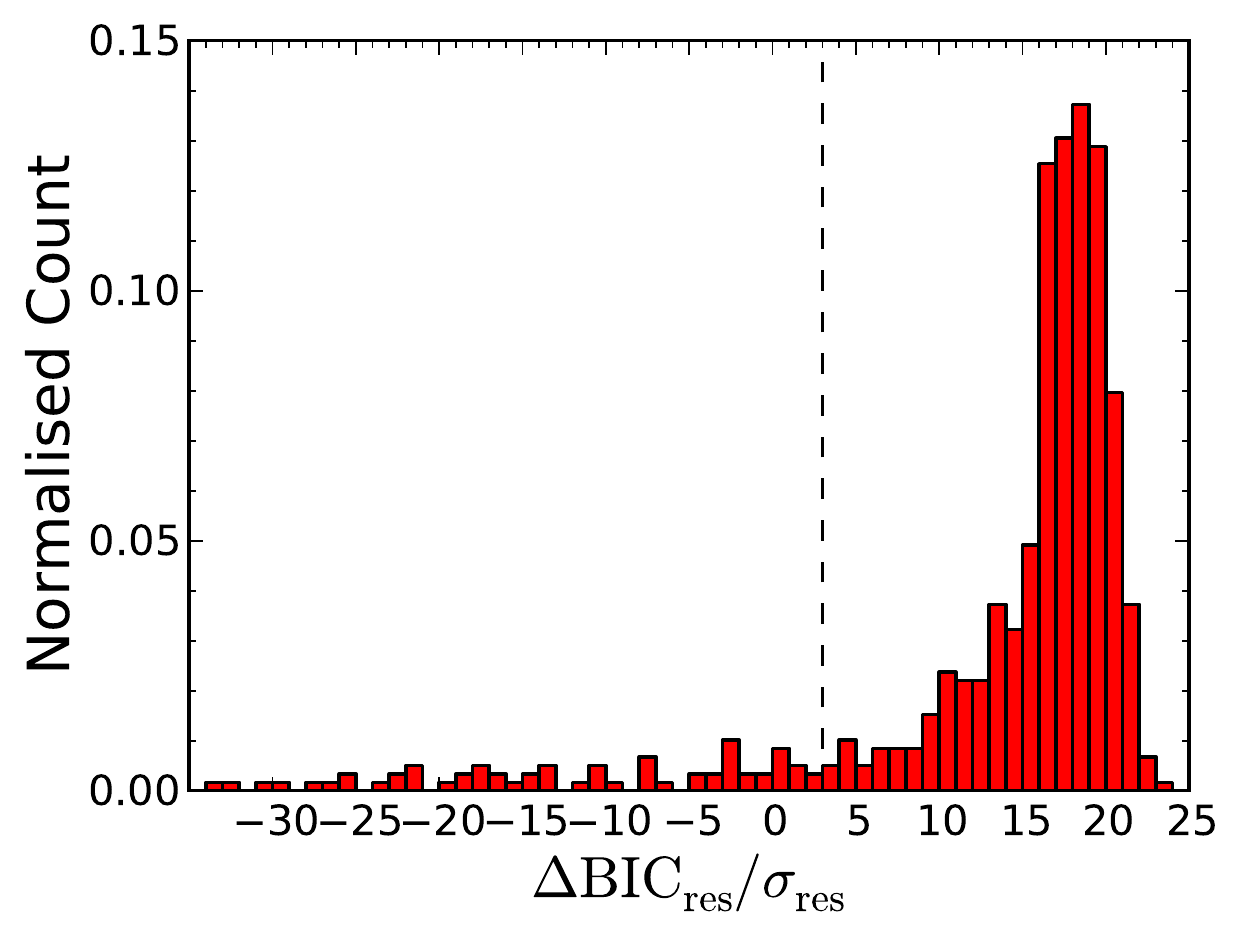}
\end{center}
\caption{Distribution of $\Delta{\rm{BIC}_{res}}$ (for comparison between \Sersic-only and \sersic + disc models) plotted in bins of width $\sigma_{\rm{res}}$, and normalised to the total Coma sample size ($N=571$). A dashed line is given at the $3\sigma_{\rm{res}}$, below which the \Sersic-only model is preferred.}
\label{BICplt}
\end{figure}

The raw (i.e. not k- or extinction-corrected) 2-component GALFIT results require vetting to select a sample for which an `archetypal' \sersic + exponential model is an appropriate interpretation of the underlying galaxy structure. A logical filter was applied to identify and remove unstable, \Sersic, and atypical 2-component fits, then to flag potential anomalous archetypal 2-component fits (in that order).

Unstable fits are characterised by a high value of $\chi^2$, indicating poor goodness-of-fit. Galaxies with significant contamination, or strong asymmetrical features also cannot be fit reliably, but may yield artificially low values of $\chi^2$ (e.g. due to overzealous masking). To remove all unstable fits, more robust testing than an upper limit to permitted values of $\chi^2$ is required. 

First, science thumbnails were tested for contamination by neighbouring SExtractor-detected sources. The fraction of masked pixels within an ellipse with semimajor axis, $a$, and axis ratio, $q$, is calculated as,
\begin{equation}
f_{\rm{mask}}(a,q) = \frac{n_{\rm{mask}}(a,q)}{n_{\rm{pix}}(a,q)},
\end{equation}
where $n_{\rm{pix}}$ is the total number of pixels contained in the ellipse, and $n_{\rm{mask}}$ is the number of masked pixels. If the fraction of masked pixels within the target ellipse (see Section \ref{prep}), $f_{\rm{mask}}(a_{\rm{target}},q_{\rm{target}})\geq0.4$, or if the fraction within the inner quarter of the target ellipse, $f_{\rm{mask}}(a_{\rm{target}}/2,q_{\rm{target}})\geq0.3$, then the thumbnail is considered contaminated. 

Secondly, the thumbnails were tested for asymmetric features using the asymmetry parameter, $A$ \citep{Homeier2006}. $A$ is defined as

\begin{equation}
A = \frac{1}{2}\left[\Sigma(|I-I_{180}|) - B_{\rm{corr}}\right]\frac{1}{I_t},
\end{equation} 
where $I$ is the flux in a particular pixel, $I_{180}$ is the flux in the same pixel after rotating the image by 180$^{\circ}$, and $I_t$ is the sum of flux in all pixels. $B_{\rm{corr}}$ is a correction for uncorrelated noise equivalent to calculating $\Sigma(|I-I_{180}|)$ for a empty region of sky. If a galaxy image has asymmetry (as calculated in the unmasked target ellipse), $A>0.2$, then it is removed from analysis as unstable.The galaxy is also removed if the asymmetry calculated for the \sersic+exponential model subtracted residual $A_{\rm{res}}>0.3$.

Lastly, we remove any remaining 2-component model fits with $\chi^2_{\nu}\geq1.7$ as unstable. This cut was calibrated on visual inspection of fit residuals.

To detect which galaxies are better fit by a pure \sersic model, we use the Bayesian Information Criterion (BIC, \citealp{BIC}). BIC modifies the standard $\chi^2$ assessment of goodness-of-fit to penalise the addition of unnecessary free parameters. Thus, this statistic can be used to identify galaxies for which the addition of a (disc) component does not significantly improve the model fit. The general form of BIC is

\begin{equation}
{\rm{BIC}} = \chi^2 + k\cdotp{\rm{ln}}(n)
\label{BIC_eq}
\end{equation}

where $k$ is the number of model free parameters, and $n$ is the number of independent data points. When comparing two fitting models, A and B, the fit which results in the lowest BIC is considered the preferred model. Thus, if $\Delta{\rm{BIC}} = {\rm{BIC}_A} - {\rm{BIC}_B} > 0$, model B provides a better fit than model A, regardless of any difference in number of free parameters.

A BIC test is selected over the similar Akaike Information Criterion (AIC, \citealp{AIC}) as it more strongly penalises unnecessary model parameters. Thus, the BIC is a more thorough statistic for the identification of overfit models (e.g. B+D fits to 1-component galaxies). As this work focuses on a clean sample of 2-component galaxies, ``false-negative" detections of overfitting (i.e. 2-component galaxies catagorised as 1-component) is preferable to ``false positive" detections (i.e. 1-component galaxies catagorised as 2-component). Furthermore, in contrast to an F-test, the BIC allows ``non-nested" model comparison even where simpler models {\emph{cannot}} be expressed in the form of more complex models (i.e. comparison of boxy \sersic and \sersic+ exponential models). 

Independence of data points is a core assumption of the BIC, however the individual pixels in the image thumbnails cannot be considered statistically independent. Instead, model selection must be evaluated from all independent {\emph{resolution}} elements. Following the prescription of \cite{Simard2011}, the number of pixels, $n_{\rm pix}$, in Equation \ref{BIC_eq} is substituted for the number of resolution elements, $n_{\rm res} = n/A_{\rm psf}$, where $A_{\rm{psf}}$ is the size of the resolution element in pixels. 

For consistency, the fitting chi-squared must also be evaluated across independent resolution elements. However, the identification of which pixels contribute to each resolution element is a non-trivial problem. Instead, we approximate the resolution-chi-squared as, $\chi^2_{\rm res} \approx \chi^2{\rm pix}/A_{\rm psf}$. This approach is equivalent to evaluation of $\chi^2_{\rm res}$ via summation of the {\emph{average contributions}} to $\chi^2$ of each resolution element. The resolution-modified BIC, $\rm{BIC}_{\rm{res}}$, is thus:

\begin{equation}
{\rm{BIC}}_{\rm{res}} = \frac{\chi^2}{A_{\rm{psf}}} + k\cdotp{\rm{ln}}\left(\frac{n_{\rm{pix}}}{A_{\rm{psf}}}\right).
\end{equation}

\begin{figure}
\begin{center}
\includegraphics[width=\linewidth]{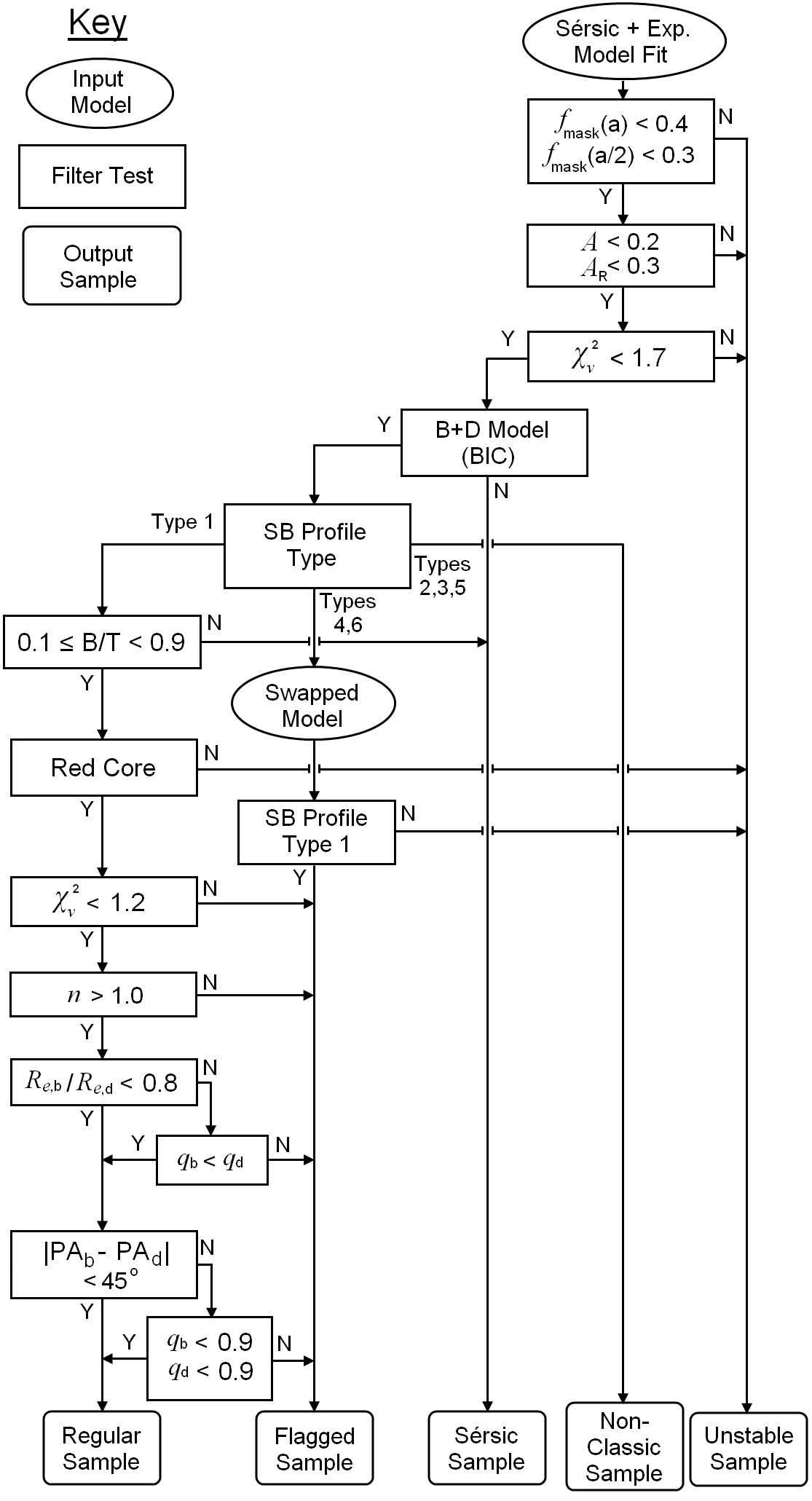}
\end{center}
\caption{Flow chart for sample filtering following \sersic + exponential fitting. For profile type definitions, refer to \citet{Allen2006}.}
\label{flowfilt}
\end{figure}

Since measurement of $A_{\rm{psf}}$ has an associated error ($\sim5\%$), the use of ${\rm{BIC}_{res}}$ introduces uncertainty to the value of $\Delta{\rm{BIC}}$. This error, $\sigma_{\rm{res}}$, is estimated via monte carlo simulation. Thus, when comparing models we select the simpler model unless $\Delta{\rm{BIC}_{res}} >3\sigma_{\rm{res}}$. This is illustrated in Figure \ref{BICplt}, where the distribution of $\Delta{\rm{BIC}_{res}}$ is plotted in $\sigma_{\rm{res}}$ bins. The majority of galaxies possess a $\Delta{\rm{BIC}_{res}}$ significantly higher than 0, with a long tail towards negative values. Setting the acceptance limit for 2-component galaxies at $3\sigma{\rm{res}}$ thus only affects a small minority of galaxies, but ensures an analysis sample uncontaminated by misclassified 1-component galaxies.

Model selection tests via BIC$_{\rm res}$ and pixel-wise BIC (BIC$_{\rm pix}$) were compared to visual model selection via inspection of 1D surface brightness profiles. For the majority ($\sim83\%$) of (uncontaminated, symmetric) galaxies, all three tests select the same `best-fit' model. Where the test results {\emph{do not}} agree, BIC$_{\rm pix}$ selects models which are overfit (relative to the visually-selected `best-fit'), while BIC$_{\rm res}$ selects an underfit model ($\sim10\%$ and $\sim3\%$ of cases respectively). As identification of a clean sample of well-fit bulge + disc galaxies is the primary goal of model selection in this work, the incorrect rejection of a small number of bulge + disc galaxies is vastly preferable to contamination by overfit 1-component galaxies. Thus, we use BIC$_{\rm res}$ to distinguish between the goodness-of-fit of the \sersic ($k=7$), \sersic + exponential ($k=11$), and and boxy \Sersic ($k=8$) models in this work. 

A 1D BIC test is also used during multi-band fitting to select between component gradient models. Radial colour profiles (for the image and models) are measured in elliptical annuli (in the $u$- and $i$-bands) within the unmasked target ellipse. BIC is evaluated as per Equation B3, where $n$ is the number of annuli, for $k=8$ (no gradient), $k=10$ ($\times2$; disc or bulge gradient), and $k=12$ (bulge and disc gradients). 

Surface brightness profiles for all 2-component galaxies are categorised by \cite{Allen2006} type, based on the relative dominance of each component at different radii. For this purpose, the component surface brightness profiles are calculated analytically in the $g$-band out to the edge of the target ellipse, and parametrised by which component dominates the model centre, and the number of times they cross. 

Profile Types 4 (exponential-dominated core, crosses once) and 6 (exponential-dominated core, crosses twice) are inappropriate for a ``bulge + disc" interpretation and are moved to the unsuitable sample. Galaxies with profile Types 2 (exponential-dominated core, does not cross), 3 (\Sersic-dominated core, crosses twice), and 5 (\Sersic-dominated core, never crosses) are atypical bulge + disc models, and define the ``atypical" sample. The remaining galaxies have an archetypal Type 1 profile (\Sersic-dominated core, crosses once).

It is possible to recover a fraction of the inverted (Type 4) fits by swapping the parameter values of the \sersic and exponential components and re-fitting. When the re-fit model yields a type 1 profile, the galaxy is moved to the analysis sample. However, as forced parameter swap may result in a higher value of $\chi^2_{\nu}$ ($\lesssim0.01$), galaxies recovered in this way are flagged as potentially unreliable. Such galaxies may be better fit by a double \sersic model. 

GALFIT includes a bias towards non-zero component fluxes, such that the fit to one-component galaxy would contain a non-negligible second component. A typical \sersic galaxy will be fit by a \sersic+ exponential model with B/T as low as $\sim0.9$. Thus, we remove all fits with $B/T > 0.9$ as indistinguishable from a \Sersic-only galaxy. Fits with B/T $< 0.1$ are likewise removed, as below this value the photometric uncertainty associated with the bulge component becomes very large. Galaxies removed due to B/T cuts are moved to the \sersic sample.

Radial colour profiles are measured for use during multi-band fitting (see Section \ref{colgradintro}). A two-component model is insufficient to reproduce a galaxy colour profile which includes a significant downturn towards bluer colours in the core. Thus, if the innermost colour data point (as measured from the galaxy thumbnails) is more than 0.01 mag bluer than the adjacent measurement, that galaxy is moved to the unsuitable sample.

All remaining galaxies are included in the analysis sample. These galaxies are subjected to further filtering to identify less reliable fits: A cut at \Chi$ > 1.2$ is used to flag poorer fits caused by weak dust lanes or asymmetries, and unfit spiral patterns or nuclear bars. Models in which the \sersic component follows a bar rather than the bulge are also identified by flagging `bulges' with low \sersic indicies ($n<1.0$). Bar-like discs are flagged where the exponential component is smaller ($R_{e,D}/R_{e,B}> 0.8$) and narrower ($q_D < q_D$) than the \sersic component. A large PA misalignment between the two model components (${\rm |PA_D-PA_B|} > 45^{\circ}$) also indicates a bar-like structure, so such galaxies are flagged unless one or both components are round ($q>0.9$).

\setcounter{figure}{0}
\renewcommand{\thefigure}{C\arabic{figure}}
\begin{figure}
\begin{center}
\includegraphics[width=\linewidth]{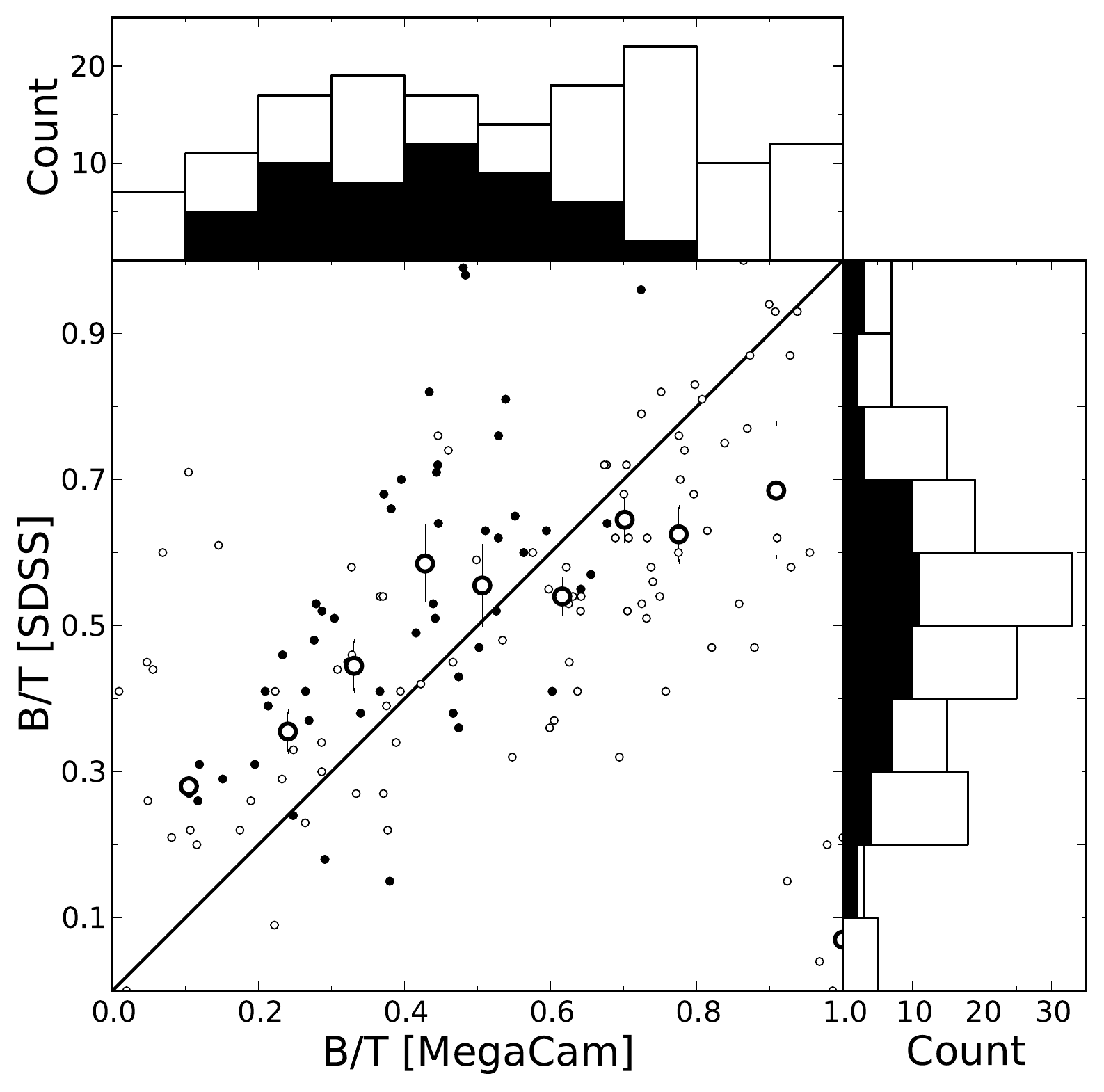}
\end{center}
\caption{Comparison of the ($g$-band) B/T for Coma cluster galaxies as measured in the present work (x-axis), and from \protect\cite{Simard2011} (y-axis). Small unfilled points (and the unfilled portions of histograms) indicate galaxies which would be removed from analysis by sample filtering. Large unfilled points with errors indicate median values (for all plotted galaxies) measured in B/T bins. A $y=x$ line is included for comparison.}
\label{Sim_plot}
\end{figure}

\setcounter{figure}{1}
\renewcommand{\thefigure}{B\arabic{figure}}
The filtering process is described diagrammatically in Figure \ref{flowfilt}. The results of applying the filter to the initial galaxy sample are given in Section \ref{filt_prop}.

\setcounter{figure}{1}
\renewcommand{\thefigure}{C\arabic{figure}}

\setcounter{table}{0}
\renewcommand{\thetable}{D\arabic{table}}
\begin{table*}
\begin{tabular}{|l|l|l|l|l|l|l|l|l|l|l|l|}
\hline
\hline
(1) & (2) &  (3) & (4) & (5) & (6) & (7) & (8) & (9) & (10) & (11) &  (12) \\
(13) & (14) & (15) & (16) & (17) & (18) & (19) &  (20) & (21) & (22) & (23) & (24) \\
(25) & (26) & (27) & (28) & (29) & (30) & (31) & (32) & (33) &  (34) & (35) & (36) \\
\hline
\hline
1237665427552927881&194.8755&28.7001&44.19&0.024&-17.483&0.012&-15.777&0.146&-17.128&0.021&-17.403\\0.037&0.215&0.023&2.628&0.048&2.121&0.026&0.843&0.006&1.066&0.224&2.614\\0.034&10.000&0.929&0.826&0.034&0.839&0.014&20.0&0.98&0.94&3&U4 \\ \hline
1237665427552993436&195.0183&28.6035&38.22&0.023&-19.280&0.003&-16.369&0.114&-19.000&0.005&-19.092\\0.010&0.075&0.007&10.058&0.057&2.150&0.009&0.783&0.002&1.310&0.217&8.422\\0.025&1.375&0.120&0.571&0.023&0.796&0.003&154.6&1.07&0.91&1&S7 \\ \hline
1237665427552993478&195.0946&28.5745&37.01&0.022&-18.446&0.007&-16.778&0.072&-17.920&0.014&-18.245\\0.021&0.251&0.013&9.734&0.109&2.589&0.020&0.801&0.003&3.232&0.323&8.608\\0.072&2.013&0.101&0.575&0.013&0.947&0.011&60.1&0.98&0.95&1&A11\\ \hline
1237665427553124587&195.4493&28.6602&48.74&0.029&-17.415&0.015&-16.184&0.129&-16.979&0.033&-17.405\\0.047&0.315&0.026&4.412&0.102&1.579&0.024&0.730&0.006&2.076&0.379&6.384\\0.127&1.046&0.083&0.779&0.028&0.685&0.018&2.0&0.96&0.95&1&S6\\ \hline
1237665427553189967&195.4972&28.7095&52.59&0.020&-17.171&0.017&-17.094&0.049&-14.386&0.129&-17.180\\0.046&0.921&0.009&2.525&0.065&2.172&0.036&0.665&0.006&2.673&0.173&1.490\\0.108&2.153&0.065&0.743&0.011&0.261&0.023&0.3&0.95&0.94&5&S6\\
\hline
\hline
\end{tabular}
\caption{The structural and photometric parameters of \sersic and \sersic + exponential models fits ($g$-band) for the entire galaxy sample. The column headings are described in Table \ref{res1_key}. This table displays the first five data rows only; the complete version will be made available online.}
\label{result_1}
\end{table*}

\setcounter{table}{2}
\begin{table*}
\begin{tabular}{|l|l|l|l|l|l|l|l|l|l|l|l|}
\hline
\hline
(i) & (ii) &  (iii) & (iv) & (v) & (vi) & (vii) & (viii) & (ix) & (x) & (xi) &  (xii) \\
(xiii) & (xiv) & (xv) & (xvi) & (xvii) & (xviii) & (xix) &  (xx) & (xxi) & (xxii) & (xxiii) & (xxiv) \\
(xxv) & (xxvi) & (xxvii) & (xxviii) & (xxix) & (xxx) & (xxxi) & (xxxii) & (xxxiii) & (xxxiv) &  (xxxv) & (xxxvi)  \\
\hline
\hline
1237665427552927881&-14.034&0.173&-15.624&0.051&-15.850&0.052&0.181&0.025&1.066&0.269&1.558\\0.019&0.917&-16.936&0.133&-18.156&0.016&-18.461&0.035&0.236&0.022&1.066&0.202\\1.558&0.021&0.930&-0.193&0.133&-0.056&0.018&-0.338&0.151&-0.078&0.019&N\\ \hline
1237665427552993436&-14.697&0.130&-17.473&0.013&-17.554&0.015&0.065&0.007&1.310&0.251&5.019\\0.015&0.933&-17.450&0.107&-20.012&0.004&-20.110&0.010&0.079&0.007&1.310&0.203\\5.019&0.015&0.935&-0.047&0.073&-0.048&0.004&-0.072&0.079&-0.066&0.004&N\\ \hline
1237665427552993478&-15.282&0.069&-16.377&0.034&-16.715&0.031&0.260&0.014&3.232&0.310&5.130\\0.044&0.943&-17.823&0.068&-18.873&0.012&-19.222&0.021&0.267&0.012&3.232&0.302\\5.130&0.044&0.921&0.064&0.148&0.029&0.007&0.017&0.111&-0.023&0.007&N\\ \hline
1237665427553124587&-14.727&0.135&-15.612&0.064&-16.010&0.061&0.302&0.029&2.076&0.397&3.805\\0.074&0.969&-17.149&0.122&-17.856&0.026&-18.311&0.045&0.333&0.025&2.076&0.357\\3.805&0.078&0.987&-0.110&0.141&-0.048&0.011&-0.167&0.142&-0.137&0.009&N\\ \hline
1237665427553189967&-15.692&0.047&-11.933&0.284&-15.726&0.046&0.969&0.008&2.673&0.164&0.888\\0.071&0.951&-18.065&0.049&-15.467&0.093&-18.160&0.046&0.912&0.008&2.673&0.175\\0.888&0.063&0.936&-0.351&0.167&-0.140&0.018&-0.442&0.195&-0.192&0.012&N\\
\hline
\hline
\end{tabular}
\caption{The multiband fitting parameters of \sersic + exponential models fits ($u$- and $i$-bands) for the entire galaxy sample. Any deviations from the $g$-band structures (see Table \ref{result_1}) are included (i.e. where internal component gradients are necessary). The column headings are described in Table \ref{res2_key}. This table displays the first five data rows only; the complete version will be made available online.}
\label{result_2}
\end{table*}

\section{Comparison with Other Work}
\setcounter{figure}{1}
\begin{figure}
\begin{center}
\includegraphics[width=\linewidth]{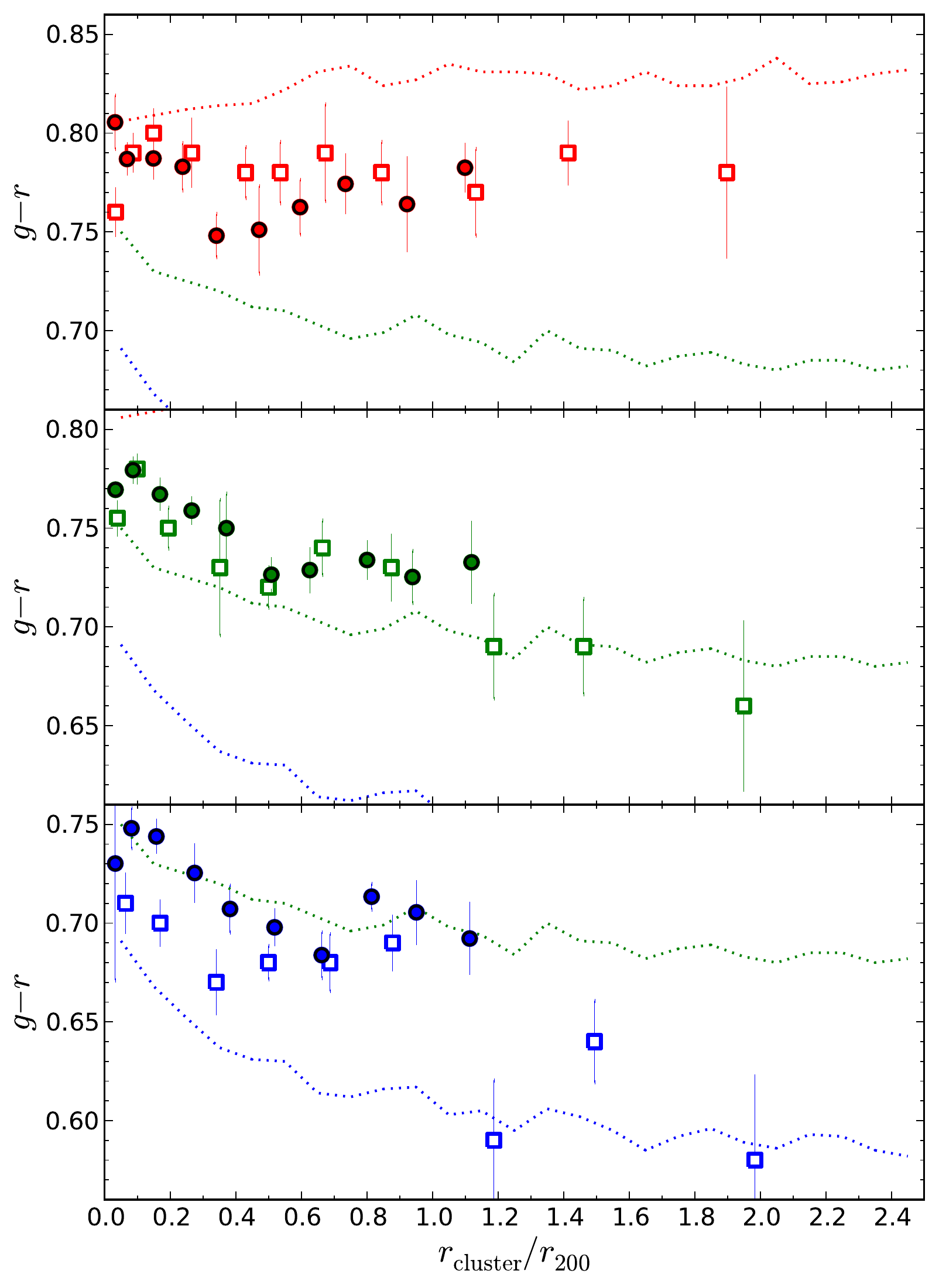}
\end{center}
\caption{Colour-(cluster) radius plots for bulge (red, upper panel), total galaxy (green, middle panel), and disc (blue, bottom panel) $g-r$ colours. Filled circular points indicate the results of the present work (converted from $g-i$). Unfilled square points indicate the data from \protect\cite{Simard2011} for Coma. The equivalent data for rich clusters from \protect\cite{Taranu2012} are included in each panel for comparison. Plotted colour data (from all three sources) are median values measured in $r_{\rm cluster}$ bins.}
\label{Tar_plot}
\end{figure}

\subsection{Comparison with HST}
\cite{Weinzirl2013} present \sersic fit parameters for 68 Coma galaxies derived from the Hubble Space Telescope F814W imaging of the ACS Coma Cluster Treasury survey \citep{Carter2008}. For the $\sim$\,50 galaxies in common with the initial sample of the present work, our \Sersic-only structural parameters ($n$, $R_e$) agree at the $\sim$\,$10\%$ level with the values reported by \citeauthor{Weinzirl2013}. A few notable outliers ($N\sim8$) exist for both parameters, resulting from differing treatment of deblending. A detailed comparison of our MegaCam-based structural measurements with those derived from HST imaging will be presented elsewhere.

\subsection{Comparison with SDSS}
In this section, the bulge + disc model parameters from the present work (hereafter `MegaCam Coma') are compared to bulge-disc decomposition of SDSS imaging for both Coma (from \citealp{Simard2011}; hereafter `SDSS Coma') and a large galaxy sample from a set of rich ($\sigma_{\rm 1D} > 800$ km s$^{-1}$) clusters (\citealp{Taranu2012}; hereafter T12). Note that the T12 data is also drawn from the \cite{Simard2011} decomposition catalogue.

For comparison with the SDSS samples, we use our entire (unfiltered) Coma sample, as neither SDSS Coma or T12 filter their galaxy sample to remove unsuitable model fits. The colour cut ($g-r > 0.5$) is removed from our sample for the same reason. The T12 selection criteria are then applied to our data (and used to select a Coma sample from \citealp{Simard2011}) to ensure equivalence of the compared samples: the magnitude limit for galaxies is thus set to $M_g < -19.3$ ($\approx M_R <-20.25$, i.e. $\sim2$ mag fainter than our sample limit). The resulting MegaCam and SDSS Coma samples have $212$ and $177$ galaxies respectively. 

The distributions of ($g$-band) B/T measured for the MegaCam and SDSS Coma samples are presented in Figure \ref{Sim_plot}. For B/T $<0.6$, both estimates are well correlated, with a significant offset (such that the SDSS B/T is systematically larger than for Megacam on average) and substantial scatter. For more bulge-dominated galaxies, SDSS B/T is weakly correlated with MegaCam B/T, and offset in the opposite direction (SDSS B/T systematically smaller than MegaCam). This apparent bimodality (as evident from the double-peaked distribution for MegaCam B/T) reflects the difference between archetypal (low B/T)  and atypical ($\equiv$ traditional giant ellipticals; high B/T) galaxy fits in our sample. 

Colour-cluster radius plots are displayed in Figure \ref{Tar_plot} for bulge, disc and total galaxy colours. As per T12, bulge data are only included if B/T $> 0.2$, and disc data are only included if B/T $< 0.8$. The $g-r$ colours are calculated for MegaCam Coma data using $(g-r)=0.67(g-i)+0.01$ for bulge and galaxy colours and $(g-r)=0.53(g-i)+0.15$ for disc colours. These conversion maps are measured empirically from SDSS aperture photometry in the appropriate colour ranges.

Coma bulges are found to be systematically bluer (from MegaCam and SDSS) than bulges in the full T12 cluster sample. This offset is less prominent in the cluster core, but increases with $r_{\rm cluster}$ due to a weakly positive radial trend in T12 and a weakly negative/flat trend for MegaCam/SDSS. On average, T12 bulges are $\sim0.04$ mag redder in $g-r$. Total galaxy colours are marginally redder on average in Coma than in T12, albeit with equivalent radial trends in all three datasets. MegaCam galaxies are $\sim0.03$ mag redder in $g-r$ than T12 on average, while SDSS galaxy colours are in agreement with both MegaCam and T12.

Discs in Coma galaxies are found to be systematically redder than T12 discs, although SDSS disc colours are consistent with T12 beyond $r_{200}$. MegaCam Coma discs are $\sim0.08$ mag redder than T12 ($\sim0.03$ mag redder than SDSS in Coma), but follow a similar negative radial colour trend. Similarly, the discs of SDSS Coma galaxies are $\sim0.05$ mag redder than T12 in $g-r$, with a noisy (negative) radial trend in colour. The discrepancy between Coma galaxy disc colours may be caused by the improved disc detection from MegaCam imaging due to increased depth relative to SDSS. Discs measured from MegaCam dominate to lower galaxy-centric radii than in SDSS (i.e. lower B/T). Hence, inner galaxy flux is traded from the (SDSS) bulge to the (MegaCam) disc. As galaxies have redder centres, this makes the MegaCam discs redder on average.

To summarise, MegaCam colours agree well with decomposition of SDSS data except for disc colours, which are systematically redder than SDSS. The bulges/discs of Coma galaxies are bluer/redder in $g-r$ on average than the same components measured from a larger cluster galaxy sample.  Discrepancy between Coma SDSS component colours and T12 (despite both using the same decomposition catalogue) implies that either the Coma cluster is not representative of the ensemble properties of the rich cluster considered, or the increase in size of the seeing disc in physical units (i.e. kpc; MegaCam Coma $<$ SDSS Coma $<$ SDSS higher $z$ clusters) causes a systematic trend in structural component colours.

\section{Fitting Results Catalogue}\label{Cat}
\setcounter{table}{1}
The fitting results for the entire Coma cluster sample ($N=571$) are presented for \Sersic-only and \sersic + exponential $g$-band models in Tables \ref{result_1}, and for (dependent) \sersic + exponential $u$- and $i$-band models in Table \ref{result_2}. Independent fitting was carried out in the $g$-band. Thus, the $g$-band structural parameters (Table \ref{result_1}) were used for $u$- and $i$-band models except where internal component gradients are included. The best-fit multi-band fitting model (see Section \ref{colgradintro}, Appendix \ref{filter}) have either no internal component colour gradients (`N'), or internal colour gradients in the bulge (`B'), disc (`D'), or both components (`BD'). This is indicated by column (xxxvi) in Table \ref{result_2}. The results of sample filtering (see Appendix \ref{filter}) are included in Table \ref{result_1}, along with some of the key parameters involved in the filtering process (asymmetry, masking fraction, $\chi^2_{\nu}$. The meaning of the filtering sample codes used in Table \ref{result_1} are summarised in Table \ref{result_3}. Erroneous parameter values (e.g. in fits which failed to converge) are set to 999 in both tables.

\begin{table}
\begin{tabular}{|l|c|l}
\hline
\hline
Column name & (n) & Description  \\
\hline
\hline
ObjID& (1) 			& SDSS DR8 Object ID\\
RA& (2) 				& Object Right Ascension [degrees]\\
Dec& (3) 				& Object Declination [degrees]\\
\multirow{2}{*}{$R_{\rm{proj}}$}& \multirow{2}{*}{(4)}&Projected distance from Coma\\
&				      	& cluster centre [arcmin]\\
$z$& (5)& Object SDSS Redshift\\
\multirow{2}{*}{$M_{g{\rm{,S}}}$}&\multirow{2}{*}{(6)} &Rest-frame \sersic magnitude\\
& & ($g$-band \sersic model)\\
$e_{M_{g{\rm{,S}}}}$ & (7) & Uncertainty in $M_{g{\rm{,S}}}$ \\
\multirow{2}{*}{$M_{g{\rm{,B}}}$}& \multirow{2}{*}{(8)} 	& Rest-frame bulge magnitude\\
& & ($g$-band B+D model)\\
$e_{M_{g{\rm{,B}}}}$ & (9) & Uncertainty in $M_{g{\rm{,B}}}$ \\
\multirow{2}{*}{$M_{g{\rm{,D}}}$}&\multirow{2}{*}{(10)} & Rest-frame disc magnitude\\
& & ($g$-band B+D model)\\
$e_{M_{g{\rm{,D}}}}$ & (11) & Uncertainty in $M_{g{\rm{,D}}}$ \\
\multirow{2}{*}{$M_{g{\rm{,T}}}$}& \multirow{2}{*}{(12)} &Rest-frame total magnitude\\
& &($g$-band B+D model)\\
$e_{M_{g{\rm{,T}}}}$ & (13) & Uncertainty in $M_{g{\rm{,T}}}$\\
${\rm{B/T}}_g$& (14) 	& $g$-band bulge fraction (B+D model)\\
$e_{{\rm{B/T}}_g}$ & (15) & Uncertainty in ${\rm{B/T}}_g$ \\
\multirow{2}{*}{$R_{e{\rm{,S}},g}$}& \multirow{2}{*}{(16)} &Effective \sersic half-light radius\\
& 					&($g$-band \sersic model) [arcsec]\\
$e_{R_{e{\rm,S}},g}$ & (17) & Uncertainty in $R_{e{\rm{,S}},g}$ \\
$n_{{\rm{S}},g}$& (18) &\sersic index ($g$-band \sersic model)\\
$e_{n_{{\rm{S}},g}}$ & (19) & Uncertainty in $n_{{\rm{S}},g}$ \\
\multirow{2}{*}{$q_{{\rm{S}},g}$}& \multirow{2}{*}{(20)} &  \sersic axis ratio\\
& & ($b/a$,$g$-band \sersic model)\\
$e_{q_{{\rm{S}},g}}$ & (21) & Uncertainty in $q_{{\rm{S}},g}$  \\
\multirow{2}{*}{$R_{e{\rm{,B}},g}$}& \multirow{2}{*}{(22)} & Effective bulge half-light radius\\
& & ($g$-band B+D model) [arcsec]\\
$e_{R_{e{\rm{,B}},g}}$ & (23) & Uncertainty in $R_{e{\rm{,B}},g}$ \\
\multirow{2}{*}{$R_{e{\rm{,D}},g}$}& \multirow{2}{*}{(24)} &  Effective disc half-light radius\\
& & ($g$-band B+D model) [arcsec]\\
$e_{R_{e{\rm{,D}},g}}$ & (25) & Uncertainty in $R_{e{\rm{,D}},g}$\\
$n_{\rm{B}}$& (26) & Bulge \sersic index (B+D model)\\
$e_{n_{\rm{B}}}$ & (27) & Uncertainty in $n_{\rm{B}}$ \\
$q_{\rm{B}}$& (28) & Bulge axis ratio ($b/a$, B+D model)\\
$e_{q_{\rm{B}}}$ & (29) & Uncertainty in $q_{\rm{B}}$ \\
$q_{\rm{D}}$& (30) & Disc axis ratio ($b/a$, B+D model)\\
$e_{q_{\rm{D}}}$ & (31) & Uncertainty in $q_{\rm{D}}$ \\
\multirow{2}{*}{$\Delta({\rm{PA}})$}& \multirow{2}{*}{(32)} 	&Absolute position angle difference \\
& 						&between components [Degrees]\\
$\chi^2_{{\nu,{\rm{s}}},g}$& (33) & Reduced $\chi^2$ ($g$-band, \sersic model)\\
$\chi^2_{{\nu},g}$& (34) & Reduced $\chi^2$ ($g$-band, B+D model)\\
\multirow{2}{*}{ProfType}& \multirow{2}{*}{(35)} & Major axis surface brightness\\
& &profile type (see \citealp{Allen2006})\\
FiltSamp& (36) &  Filtering sample (Table \ref{result_3})\\
\hline
\hline
\end{tabular}
\caption{This table describes the column headings (1-36) for Table \ref{result_1}. Fitting results are presented for \Sersic-only and bulge + disc (``B+D") models in the $g$-band.}
\label{res1_key}
\end{table}
\setcounter{table}{3}
\begin{table}
\begin{tabular}{|l|l|l}
\hline
\hline
Column name & (n) & Description  \\
\hline
\hline
ObjID& (i) & SDSS DR8 Object ID\\
$M_{u{\rm{,B}}}$& (ii) 		&Rest frame bulge magnitude ($u$-band)\\
$e_{M_{u{\rm{,B}}}}$ & (iii) 	&Uncertainty in $M_{u{\rm{,B}}}$ \\
$M_{u{\rm{,D}}}$& (iv) 		&Rest frame disc magnitude ($u$-band)\\
$e_{M_{u{\rm{,D}}}}$ & (v) 	&Uncertainty in $M_{u{\rm{,D}}}$ \\
$M_{u{\rm{,T}}}$& (vi) 		&Rest frame total magnitude ($u$-band)\\
$e_{M_{u{\rm{,T}}}}$ & (vii) 	&Uncertainty in$M_{u{\rm{,T}}}$\\
${\rm{B/T}}_u$& (viii) 		&$u$-band bulge fraction\\
$e_{{\rm{B/T}}_u}$ & (ix) 	&Uncertainty in ${\rm{B/T}}_u$ \\
\multirow{2}{*}{$R_{e{\rm{,B}},u}$}& \multirow{2}{*}{(x)} 	&Effective bulge half-light radius\\
& &($u$-band) [arcsec]\\
$e_{R_{e{\rm{,B}},u}}$ & (xi) & Uncertainty in $R_{e{\rm{,B}},u}$ \\
\multirow{2}{*}{$R_{e{\rm{,D}},u}$} &\multirow{2}{*}{(xii)} & Effective disc half-light radius\\
& & ($u$-band) [arcsec]\\
$e_{R_{{\rm{,D}},u}}$ & (xiii) & Uncertainty in $R_{e{\rm{,D}},u}$ \\
$\chi^2_{{\nu},u}$& (xiv) & Reduced $\chi^2$ ($u$-band)\\
$M_{i{\rm{,B}}}$& (xv) 		&Rest frame bulge magnitude ($i$-band)\\
$e_{M_{i{\rm{,B}}}}$ & (xvi) 	&Uncertainty in $M_{i{\rm{,B}}}$ \\
$M_{i{\rm{,D}}}$& (xvii) 		&Rest frame disc magnitude ($i$-band)\\
$e_{M_{i{\rm{,D}}}}$ & (xviii) 	&Uncertainty in $M_{i{\rm{,D}}}$ \\
$M_{i{\rm{,T}}}$& (xix) 		&Rest frame total magnitude ($i$-band)\\
$e_{M_{i{\rm{,T}}}}$ & (xx) 	&Uncertainty in$M_{i{\rm{,T}}}$\\
${\rm{B/T}}_i$& (xxi) &  $i$-band bulge fraction \\
$e_{{\rm{B/T}}_i}$ & (xxii) & Uncertainty in ${\rm{B/T}}_i$ \\
\multirow{2}{*}{$R_{e{\rm{,B}},i}$}& \multirow{2}{*}{(xxiii)} & Effective bulge half-light radius\\
&& ($i$-band) [arcsec]\\
$e_{R_{e{\rm{,B}},i}}$ & (xxiv) & Uncertainty in $R_{e{\rm{,B}},i}$ \\
\multirow{2}{*}{$R_{e{\rm{,D}},i}$} &\multirow{2}{*}{(xxv)} & Effective disc half-light radius\\
& & ($i$-band) [arcsec]\\
$e_{R_{{\rm{,D}},i}}$ & (xxvi) & Uncertainty in $R_{e{\rm{,D}},i}$ \\
$\chi^2_{{\nu},i}$& (xxvii) & Reduced $\chi^2$ ($i$-band)\\
$\Delta(u-g)$& (xxviii) &  Radial gradient in galaxy ($u-g$) \\
$e_{\Delta(u-g)}$ & (xxix) & Uncertainty in $\Delta(u-g)$ \\
$\Delta(u-g)_{\rm{m}}$& (xxx) &  Radial gradient in model ($u-g$)\\
$e_{\Delta(u-g)_{\rm{m}}}$ & (xxxi) & Uncertainty in $\Delta(u-g)_{\rm{m}}$ \\
$\Delta(g-i)$& (xxxii) &  Radial gradient in galaxy ($g-i$)\\
$e_{\Delta(g-i)}$ & (xxxiii) & Uncertainty in $\Delta(g-i)$ \\
$\Delta(g-i)_{\rm{m}}$& (xxxiv) &  Radial gradient in model ($g-i$)\\ 
$e_{\Delta(g-i)_{\rm{m}}}$ & (xxxv) & Uncertainty in $\Delta(g-i)_{\rm{m}}$ \\
MultiType & (xxxvi) & Best-fit multi-band model \\
\hline
\hline
\end{tabular}
\caption{This table describes the column headings (i-xxxvi) for Table \ref{result_2}. Results are presented for bulge + disc models in the $u$- and $i$-bands.}
\label{res2_key}
\end{table}

\begin{table}
\begin{tabular}{|c|c|l}
\hline
\hline
Sample & Subsample & Reason for filtering  \\
\hline
\hline
\multirow{5}{*}{Unsuitable (U)}& 1 & Masking fraction \\
& 2 & Asymmetry \\
& 3 & High $\chi^2_{\nu}$\\
& 4 & SB profile type\\
& 5 & Blue core\\
\hline
\multirow{2}{*}{\sersic (S)} & 6 & BIC\\
& 7 & B/T \\
\hline
\multirow{4}{*}{Analysis (A)} & 8 & Flagged ($\chi^2$)\\
& 9 & Flagged (Bar-like structure)\\
& 10 & Flagged (Forced swap)\\
& 11 & Regular fit \\
\hline
\hline
\end{tabular}
\caption{Filtering sample/subsample codes for Table \ref{result_1}.}
\label{result_3}
\end{table}

\label{lastpage}
\end{document}